\documentclass[reprint, aps, prx, amsmath, amssymb, longbibliography, superscriptaddress]{revtex4-2}

\usepackage{graphicx}
\usepackage{epsfig}
\usepackage{mathrsfs,esint}
\usepackage{physics}
\usepackage{relsize}
\usepackage{bbold}
\usepackage{dsfont}
\usepackage{comment}
\usepackage{svg}
\usepackage{enumitem}

\usepackage[caption=false]{subfig}

\usepackage[bookmarks=true,
   colorlinks=true,
   linkcolor=blue,
   urlcolor=blue,
   citecolor=blue,
   bookmarks=true,
   hyperindex=true
]{hyperref}

\DeclareMathOperator{\Zn}{\mathbb{Z}_n}
\DeclareMathOperator{\Zthree}{\mathbb{Z}_3}
\DeclareMathOperator{\Ztwo}{\mathbb{Z}_2}

\DeclareMathOperator{\sgn}{sign}

\begin{document}

\title{From coupled \texorpdfstring{$\Zthree$}{Z3} Rabi models to the \texorpdfstring{$\Zthree$}{Z3} Potts model}

\author{Anatoliy I. Lotkov}
\affiliation{Department of Physics, University of Basel, Klingelbergstrasse 82, 4056 Basel, Switzerland}

\author{Valerii K. Kozin}
\affiliation{Department of Physics, University of Basel, Klingelbergstrasse 82, 4056 Basel, Switzerland}

\author{Denis V. Kurlov}
\affiliation{Department of Physics, University of Basel, Klingelbergstrasse 82, 4056 Basel, Switzerland}

\author{Jelena Klinovaja}
\affiliation{Department of Physics, University of Basel, Klingelbergstrasse 82, 4056 Basel, Switzerland}

\author{Daniel Loss} 
\affiliation{Department of Physics, University of Basel, Klingelbergstrasse 82, 4056 Basel, Switzerland}
\affiliation{Physics Department, King Fahd University of Petroleum and Minerals, 31261, Dhahran, Saudi Arabia}
\affiliation{Quantum Center, KFUPM, Dhahran, Saudi Arabia}
\affiliation{RDIA Chair in Quantum Computing}
\date{\today}


\begin{abstract}
We study $\Zthree$-symmetric Rabi model that describes a three-level system coupled to two bosonic modes. We derive a mapping of the two-mode $\Zthree$ Rabi model onto a qubit-boson ring. This mapping allows us to formulate a realistic implementation of the $\Zthree$ Rabi model based on superconducting qubits. It also provides context for the previously proposed optomechanical implementation of the $\Zthree$ Rabi model. In addition, we propose a physical implementation of the $\Zthree$ Potts model via a coupled chain of $\Zthree$ Rabi models.
\end{abstract}

\maketitle

\section{Introduction}
The field of quantum simulations has entered a mature phase in which a variety of hardware platforms can reproduce many-body Hamiltonians that are analytically intractable. Trapped-ion chains have been used to implement long-range Ising models and lattice gauge theories with tens of qubits \cite{tan_domainwall_2021, kyprianidis_observation_2021,brydges_probing_2019,zhang_observation_2017,martinez_realtime_2016,meth_simulating_2025}; neutral-atom Rydberg arrays routinely emulate spin models on two-dimensional lattices \cite{madjarov_highfidelity_2020,semeghini_probing_2021,scholl_quantum_2021,manovitz_quantum_2025,bernien_probing_2017}; and semiconductor spin-qubit platforms have realized Fermi-Hubbard and other models at the few-plaquette level \cite{dehollain_nagaoka_2020, hensgens_quantum_2017, kiczynski_engineering_2022, vandiepen_quantum_2021, wang_experimental_2022, wang_probing_2023}. These successes show that controlled programmable devices can already access a wide variety of interesting physics, such as symmetry breaking, topological order, and non-equilibrium critical phenomena that lie well beyond classical reach. One of the interesting research directions is to simulate models with larger discrete symmetries, e.g., $\Zthree$ symmetry, as they often exhibit richer physics than the models with $\Ztwo$ symmetry.

Recent advances in circuit quantum electrodynamics (cQED) provide a complementary route to this goal, focusing on light–matter systems where photons mediate interactions between superconducting qubits~\cite{kurpiers_deterministic_2018,vanloo_photonmediated_2013,astafiev_resonance_2010}. In cQED, the quantum Rabi model \cite{braumuller_analog_2017, braak_integrability_2011,hwang_quantum_2015, chen_shortcuts_2021} -- which describes a two-level atom coupled to a bosonic mode -- serves as the canonical building block and possesses a discrete $\Ztwo $ (parity) symmetry. In the ultrastrong coupling regime, a chain of coupled Rabi models effectively realizes the Ising model~\cite{hwang_largescale_2013}. There, photon-induced interactions generate nearly degenerate “cat” ground states with macroscopic qubit–cavity entanglement. Extending this connection between discrete symmetries and entangled phases from $\Ztwo$ to $\Zthree$ \cite{albert_quantum_2012, zhang_z_n_2014, sedov_chiral_2020, kozin_quantum_2024} naturally leads to the paradigmatic three-state Potts model and promises richer symmetry-breaking behavior.
Recently, there has been significant interest in the $\Zthree$ quantum Rabi model from both theoretical and experimental perspectives. 
The generalization of Albert et al. of the original ($\Ztwo$) Rabi Hamiltonian to an $N$-state atom  \cite{albert_quantum_2012} established the formal framework for studying generalized Rabi models with higher discrete symmetries. Subsequently, the first proposal for the experimental realization of the $\Zthree$ Rabi model based on the optomechanical system was discussed in~\cite{sedov_chiral_2020}. Interestingly, it was shown that in the $\Zthree$ Rabi model, the superradiant transition is first-order within the mean-field approach, unlike in the $\Ztwo$ Rabi model, where it is second-order~\cite{sedov_chiral_2020}. A recent proposal has shown how a network of Josephson junctions can implement the $\Zthree$ Potts model using tunable inductive elements \cite{wauters_engineering_2024}. Here we introduce a tunable model that connects the $\Zthree$ Rabi model with the $\Zthree$ Potts model. By varying the coupling parameter, the system evolves from a few-level regime into one that spontaneously breaks the threefold rotational symmetry. 

The remainder of the paper is organized as follows. In Sec.~\ref{sec:z3-rabi}, we give a brief overview of the $\Zthree$ Rabi model and derive the mapping of the $\Zthree$ Rabi model onto a certain qubit-bosonic model. We then illustrate how this qubit-bosonic model can be realized by superconducting-qubit and optomechanical platforms. In Sec.~\ref{sec:potts-model}, we show how the $\Zthree$ Potts model can be simulated by a chain of the $\Zthree$ Rabi models. We also illustrate it with superconducting and optomechanical implementations. At the end of Sec.~\ref{sec:potts-model}, we briefly discuss the chiral $\Zthree$ Potts model and the possibility of its implementation based on a chain of the $\Zthree$ Rabi models. Finally, in Sec.~\ref{sec:conclusion}, we conclude. Some additional details are covered in the Apps.~\ref{app:arbitrary-interaction-summary}-\ref{app:disorder}.

\section{\texorpdfstring{$\Zthree$}{Z\_3} Rabi model.}
\label{sec:z3-rabi}

\subsection{Theoretical description}
\label{sec:rabi-theory}

The central topic of this paper is the $\Zthree$-symmetric Rabi and Potts models, as well as the connection between them.

In this section, we discuss a $\Zthree$-symmetric model of two bosonic modes coupled to a three-level system (qutrit). The model generalizes the usual Rabi model to the case of multiple bosonic modes and higher discrete symmetry. In the following, we refer to this model as the $\Zthree$ Rabi model. For a more detailed discussion of this model and related ones, we refer the reader to our companion paper \cite{lotkov_cat_2025b}.
The Hamiltonian reads:
\begin{equation}
\label{eq:2-mode-z3-Rabi}
\begin{aligned}
     H_{\text{R}} = &\hbar \Omega_{\text{R}} (\hat a_1^{\dagger} \hat a_1 + \hat a_2^{\dagger} \hat a_2) + B (e^{i\phi} Z + e^{-i\phi} Z^{\dagger}) \\
    &- \lambda (\hat a_1 + \hat a_2^{\dagger}) X - \lambda (\hat a_1^{\dagger} + \hat a_2) X^{\dagger},
\end{aligned}
\end{equation}
where $\hat a_i^{\dagger}\  (\hat a_i)$ with $i = 1,2$ are boson creation (annihilation) operators \footnote{Note that we use hats to distinguish the boson operators, $\hat a,\, \hat a^{\dagger},\, \hat x, \,\hat p$.}, $X$ and $Z$ are $3\times 3 $ matrices generalizing the Pauli matrices (usually referred to as the shift and clock matrices, respectively). The parameters have the following physical meaning: $\Omega_{\text{R}}$ is the frequency of the two bosonic modes, $B$ and $\phi$ define the energy levels of the qutrit, $\lambda$ is the coupling strength between the bosonic modes and the qutrit, and $\hbar$ is the Planck constant.
Let us note in passing that in principle one can have two inequivalent ways to write a two-mode $\Zthree$ Rabi model; see the Appendix of the companion paper~\cite{lotkov_cat_2025b}. We choose the specific form of the Hamiltonian~(\ref{eq:2-mode-z3-Rabi}) because it can be naturally mapped onto a realistic physical system, as we discuss below in Sect.~\ref{sec:physical-implementation}.

The $\Zthree$ shift $X$ and clock $Z$ matrices are commonly used to describe qutrit systems. They are defined as
\begin{equation}
\label{eq:shift-clock-matreces}
    X = \begin{pmatrix} 
         0 & 0 & 1 \\
         1 & 0 & 0 \\
         0 & 1 & 0 
        \end{pmatrix}, \quad
    Z = \begin{pmatrix}
         1 & 0 & 0 \\
         0 & \omega & 0 \\
         0 & 0 & \omega^2
        \end{pmatrix},
      \end{equation}
where $\omega = \exp[2\pi i/3]$ is the principal cubic root of unity. The clock and shift matrices are unitary but not Hermitian \footnote{The $X$ and $Z$ matrices allow us to generalize the qubit Pauli group $\mathrm{G}_2 = \langle \pm i^j \omega_z^k \omega_x^l\rangle_{j,k,l = 0}^1$ to the {\it qutrit} Pauli group $\mathrm{G}_3 = \langle \pm \omega^j Z^k X^l\rangle_{j,k,l= 0}^{2}$.
The $\Zthree$ symmetry group in a qutrit degree of freedom appears naturally as a part of the qutrit Pauli group $\mathrm{G}_3$. Strictly speaking, there are two $\Zthree$ subgroups in the $\mathrm{G}_3$ group: these are generated by the powers of the matrix $X$ and $Z$. While both subgroups are conjugate to each other, here we choose the subgroup generated by the $Z$ matrix, $\Zthree = \langle Z^k \rangle_{k=0}^2$.
}.

The boson part of the $\Zthree$ symmetry is simply a residue of the bosonic $\mathrm{U(1)}$ symmetry, while its qutrit part is generated by the $Z$ matrix. As a result, the $\Zthree$ symmetry generator has the following form:
\begin{equation}
     \mathcal{P}_{\text{R}} = \exp\left[\frac{2\pi i}{3}(\hat a_1^{\dagger}  \hat a_1 -  \hat a_2^{\dagger}  \hat a_2)\right] Z,
\end{equation}
where the minus sign ensures that the operator $\mathcal{P}_{\text{R}}$ commutes with the Hamiltonian~$H_{\text{R}}$.
In the following, we provide some additional details about the $\Zthree$ Rabi model that will be useful for our purposes. For a more thorough discussion, we refer to the companion paper \cite{lotkov_cat_2025b}.

One of the main results in the present paper is an explicit construction of the $\Zthree$ Potts model based on a chain of coupled $\Zthree$ Rabi models, described by Eq.~(\ref{eq:2-mode-z3-Rabi}) [see Sec.~\ref{sec:potts-model}].
One of the key ingredients in this construction is the strong boson-qutrit interaction. 
In particular, we are interested in what is generally called the {\it extreme coupling} regime in the Rabi model literature $\lambda \gg \hbar \Omega_{\text{R}}$. In addition, we require the limit of low magnetic field $\hbar \Omega_{\text{R}} \gg  B$. This coupling regime leads to the full entanglement between the boson and qutrit degrees of freedom. Specifically, the three lowest eigenstates turn out to be the $\Zthree$ cat states \cite{lotkov_cat_2025b}:
\begin{equation}
\label{eq:three-cat-states}
    |\psi_k\rangle = \frac{1}{\sqrt{3}}\sum\limits_{l=0}^2 \omega^{lk}|\omega^{-l} \lambda/\hbar \Omega_{\text{R}}\rangle_{\hat a_1}|\omega^l \lambda/\hbar \Omega_{\text{R}}\rangle_{\hat a_2} |\omega^{-l} \rangle,
\end{equation}
where the subscripts $\hat a_1$ and $\hat a_2$ denote that it is a coherent state of the first and second bosonic modes, respectively.
The states $\ket{\psi_k}$ are the $\Zthree$ generalization of the $\Ztwo$ cat states $|+\alpha\rangle|+\rangle\pm|-\alpha\rangle|-\rangle$. The latter roughly describe the ground and the first-excited states of the Quantum Rabi model~\cite{hwang_quantum_2015}. As can be seen, each of the ordinary ($\Ztwo$) cat states is a linear combination of two coherent states $\ket{\alpha}$ and $\ket{-\alpha}$ entangled with the qubit. In the $\Zthree$ case, each $|\psi_k\rangle$ is a linear combination of {\it three} coherent states $\ket{\omega^i \alpha}$, with $i = 0,1,2$. Similarly, in the two-mode $\Zthree$ Rabi model, the corresponding eigenstates $\ket{\psi_k}$ are joint cat states of the two bosonic modes and the qutrit. In the case of the $\Zthree$ Rabi model, the coherent parameter is expressed through the model's coupling constants, $\alpha = \lambda/\hbar \Omega_{\text{R}}$. In the regime in which we are interested, $\lambda \gg \hbar \Omega_{\text{R}}$, $ B \ll \hbar \Omega_{\text{R}}$, the eigenenergies of the $\Zthree$ cat states are given by:
\begin{equation}
\begin{aligned}
\label{eq:rabi-spectrum}
    \epsilon_k &= \bra{\psi_k}H_{\text{R}} \ket{\psi_k} \\
    &= 2 B e^{-3(\lambda/\hbar \Omega_{\text{R}})^2}\cos\left[\frac{2\pi k}{3} + \phi\right] - \frac{2\lambda^2}{\hbar \Omega_{\text{R}}}.
\end{aligned}
\end{equation}
In the limit $\lambda/\hbar \Omega_{\text{R}} \to \infty$, the energy splitting between the states goes to zero. However, we are mostly interested in the intermediate regime with the splitting being small but non-zero. Now that we have reviewed all the necessary details about the $\Zthree$ Rabi model~\eqref{eq:2-mode-z3-Rabi}, we proceed to map this model onto a physically realistic system.

\subsection{Qubit-boson ring}
\label{sec:physical-implementation}

Although our ultimate goal is to present a physically realistic model for the realization of the Rabi model $\Zthree$, we first discuss a construction that serves as a link between the Rabi model and specific microscopic physical implementations. 
Namely, we show how the two-mode $\Zthree$ Rabi model~\eqref{eq:2-mode-z3-Rabi} appears in a particular model describing a three-site qubit-boson (QB) ring.
This model is a central construction of this paper. Although still purely theoretical, it is more suitable for mapping to concrete physical platforms than the original $\Zthree$ Rabi model. 

Consider the QB ring with the Hamiltonian~\cite{sedov_chiral_2020}
\begin{equation}
\label{eq:physical-hamiltonian}
  \begin{aligned}
    H_{\text{QB}} &= \epsilon \sum_{j=0}^{2} \sigma_j^z
      + \hbar \Omega_{\text{QB}} \sum_{j=0}^{2} \hat b_j^{\dagger} \hat b_j
      + V_{\text{QB}},
      \\
    V_{\text{QB}} &= g \sum_{j=0}^{2}
      \bigl( \sigma_j^{+} \sigma_{j+1}^{-}
      e^{ i ( \hat x_j - \hat x_{j+1} ) } + \text{H.c.} \bigr)\\
      &=g \sum_{j=0}^{2}
      \bigl( \sigma_j^{+} \sigma_{j+1}^{-}
      e^{ i \frac{\eta}{\sqrt{2}}( \hat b_j + \hat b_j^{\dagger}- \hat b_{j+1} - \hat b_{j+1}^{\dagger} ) } + \text{H.c.} \bigr),
  \end{aligned}
\end{equation} 
where the Pauli matrix $\sigma_j^z$ corresponds to the qubit at site $j$, $\hat b_j^{\dagger}$ ($\hat b_j$) is the creation (annihilation) operator of the $j$th boson, described by a harmonic oscillator on the real line $\mathbb{R}$, $\epsilon$ defines the Zeeman splitting of the qubit energy levels; $g$ is the coupling strength, $\Omega_{\text{QB}}$ is the boson frequency. The periodic boundary condition are chosen, therefore, the position index $j$ is valued modulo 3, e.g., $j + 3 \equiv j $. We can express the dimensionless coordinate and momentum operators for each bosonic mode through the creation and annihilation operators
\begin{align}
    \hat x_j &= \frac{\eta}{\sqrt{2}}\left(\hat b_j + \hat b_j^{\dagger}\right), \\
    \hat p_j &= \frac{i}{\sqrt{2}\eta}\left(\hat b_j^{\dagger} - \hat b_j\right).
\end{align}
The creation and annihilation operators here diagonalize the Hamiltonian of the harmonic oscillator in the following form 
\begin{equation}
    H_j = \hbar \Omega_{QB}\left(\frac{\hat x^2_j}{2\eta^2} + \frac{\eta^2 \,\hat p^2_j}{2}\right),
\end{equation}
where $\eta$ is the dimensionless squeezing parameter. We prefer this over a more conventional notation since it maps naturally on the case of the superconducting circuits, i.e., $\hat x_j$ corresponds to the normalized flux $\hat \varphi_j$, $\hat p_j$ corresponds to the normalized charge $\hat n_j$. The dimensionless parameter $\eta$ is important here since it gives us an additional parameter to tune the system.

The system has a $\mathrm{U(1)}$ symmetry that conserves the $z$-component of the total spin, $ \sum_j
\sigma_j^z$, as it clearly commutes with the Hamiltonian~\eqref{eq:physical-hamiltonian}. Thus, rather than working in the full Hilbert space, we restrict ourselves to the single-excitation sector, i.e, the subspace with exactly one qubit in the $|\uparrow\rangle$ state. In other words, the single-excitation sector is given by
\begin{equation}
\label{eq:single-excitation-sector}
  \mathcal H_{\text{QB},1} = \operatorname{Span}\bigl\{ |
    \uparrow\downarrow\downarrow\rangle,
    |\downarrow\uparrow\downarrow\rangle,
    |\downarrow\downarrow\uparrow\rangle \bigr\}.
\end{equation}

In order to prove that the QB ring~\eqref{eq:physical-hamiltonian} restricted to the single-excitation sector $\mathcal{H}_{\text{QB,1}}$~\eqref{eq:single-excitation-sector} is equivalent to the $\Zthree$ Rabi model, we first apply several algebraic transformations to the Hamiltonian $H_{\text{QB}}$~\eqref{eq:physical-hamiltonian}. To start with, we use the spin-dependent momentum translation to simplify the interaction term $V_{\text{QB}}$~\eqref{eq:physical-hamiltonian}. Then, we apply the Fourier transform, which decouples one of the bosonic modes from the rest of the system. Afterwards, the resulting Hamiltonian is restricted to the single-excitation sector $\mathcal{H}_{\text{QB},1}$~\eqref{eq:single-excitation-sector}. We obtain the two-mode $\Zthree$, which is brought to canonical form with some final adjustments.

\paragraph{Spin-dependent momentum translation.}
Applying the unitary operator
$S = \prod_{j=0}^2 e^{ i \sigma_j^z \hat x_j / 2 }$ removes the exponentials in
$\hat V_{\text{QB}}$ and yields 
\begin{align}
\label{eq:after-momentum-translation}
   & S^{-1} H_{\text{QB}} S = \epsilon \sum_{j=0}^2 \sigma_j^z
      + \hbar \Omega_{\text{QB}} \sum_{j=0}^2 \hat b_j^{\dagger} \hat b_j
      + \frac{\eta^2 \hbar \Omega_{\text{QB}}}{2} \sum_{j=0}^2 \hat p_j \sigma_j^z \notag
      \\
      &\quad + \frac{3}{4}\eta^2 \hbar \Omega_{\text{QB}}
      + g \sum_{j=0}^2
        \bigl( \sigma_j^{+} \sigma_{j+1}^{-} + \sigma_j^{-} \sigma_{j+1}^{+} \bigr).
\end{align}
After transformation, the interaction term $V_{\text{QB}}$ involves only qubit degrees of freedom, at the price of an additional
spin-boson term proportional to $\hat p_j \sigma_j^z$.

\paragraph{Fourier transform.}

To proceed, we define momentum-space operators via the Fourier transform,
\begin{equation}
\hat b(k)= \tfrac{1}{\sqrt{3}}\sum_{j=0}^2e^{-2\pi i jk/3}\hat b_j,\quad S^z(k)= \sum_{j=0}^2e^{-2\pi i jk/3}\sigma_j^z,
\end{equation}
where $k = 0,1,2$. Here, we use a different normalization for the Fourier transform of the spin operators for the following to be true, $S^z(0) = \sum_j \sigma_j^z$. The Fourier transform components of the spin ladder operators $S^{\pm}(k)$ are defined analogously. We use parentheses to distinguish momentum-space operators, which depend on the wave vector $k$, from real-space operators, which carry a site index as a subscript.  In terms of these operators, Eq.~\eqref{eq:after-momentum-translation} takes the form 
\begin{align}
    \label{eq:after-fourier}
    H_{\text{FT}} = &\epsilon S^{z}(0)
      + \hbar \Omega_{\text{QB}} \sum_{k=0}^{2} \hat b^{\dagger}(k) \hat b(k)
                          \notag\\
     &+\frac{\eta^2 \hbar \Omega_{\text{QB}}}{2\sqrt{3}} \sum_{k=0}^{2} \hat p(k) S^{z}(-k) + \frac{3}{4}\eta^2 \hbar \Omega_{\text{QB}}\\
     &+ \frac{g}{3} \sum_{k=0}^{2} \bigl(e^{2\pi ik/3} S^{+}(k) S^{-}(-k) + \text{H.c.} \bigr),\notag
\end{align}
where the zero-momentum bosonic mode ($k=0$) 
couples to the $S^z(k=0)$, which is conserved.  Hence, the zero-momentum boson actually decouples from the rest of the system. We are left with only two bosonic modes, exactly as required for the $\Zthree$ Rabi model.

\paragraph{Restriction to $\mathcal H_{\text{QB},1}$.}
To restrict the Hamiltonian~\eqref{eq:after-fourier} to the single-excitation sector $\mathcal{H}_{\text{QB},1}$~\eqref{eq:single-excitation-sector}, we identify the single-excitation spin states with qutrit states, $\ket{0} \equiv \ket{\uparrow\downarrow\downarrow}$, $ \ket{1} \equiv \ket{\downarrow\uparrow\downarrow}$ and $ \ket{2} \equiv \ket{\downarrow\downarrow\uparrow} $. After discarding constant terms and the decoupled zero-momentum mode, we obtain
\begin{align}
\label{eq:rabi-wrong-basis}
    &H_{\text{QB},1} = \hbar \Omega_{\text{QB}} \sum_{k=1}^{2} \hat b^{\dagger}(k)
      \hat b(k) + g \bigl(e^{2\pi i/3} X + e^{-2\pi i/3} X^{\dagger} \bigr)
      \\
      &+ i \frac{\eta\hbar \Omega_{\text{QB}}}{\sqrt{6}} 
        \Bigl[ -\left( \hat b(1) - \hat b^{\dagger}(2) \right) Z
        - \left( \hat b(2) - \hat b^{\dagger}(1) \right) Z^{\dagger} \Bigr], \notag
\end{align}
where $X$ and $Z$ matrices act on the qutrit formed by the single-excitation sector $\mathcal{H}_{\text{QB},1}$~\eqref{eq:single-excitation-sector}.

\paragraph{Final rearrangement.}
The Hadamard transform on the spin degrees of freedom, combined with a global
bosonic $\mathrm{U(1)}$ phase rotation, brings the Hamiltonian to the canonical
$\Zthree$ two-mode Rabi form:
\begin{align}
\label{eq:effective-rabi}
    &H_{\text{R}} = \hbar \Omega_{\text{QB}}
      \bigl( \hat a^{\dagger}_1 \hat a_1 + \hat a^{\dagger}_2 \hat a_2 \bigr)
      + g \bigl(e^{2\pi i/3} Z + e^{-2\pi i/3} Z^{\dagger} \bigr) \notag
      \\
      &- \frac{\eta\hbar \Omega_{\text{QB}}}{\sqrt{6}}
      \Bigl[ ( \hat a_1 + \hat a^{\dagger}_2 ) X
        + ( \hat a_2 + \hat a^{\dagger}_1 ) X^{\dagger} \Bigr],
\end{align}
where we identified $\hat a_k = \hat b(k)$ and $\hat a_k^{\dagger} = \hat b^{\dagger}(k)$ for $k = 1,2$. 

Comparing Eq.~(\ref{eq:effective-rabi}) with Eq.~(\ref{eq:2-mode-z3-Rabi}) one explicitly identifies the parameters of the QB ring model with the corresponding parameters of the Rabi model
\begin{equation}
\label{eq:QB-RM-parameter-mapping}
  \Omega_{\text{R}} = \Omega_{\text{QB}},
  \quad B = g,
  \quad \phi = \frac{2\pi}{3},
  \quad \lambda = \frac{\eta \hbar \Omega_{\text{QB}}}{\sqrt{6}}.
\end{equation}
For future convenience, we provide an explicit matrix representation of the ``magnetic'' term,
\begin{equation}
\label{eq:superconducting-magnetic-term}
  g \bigl(e^{2\pi i/3} Z + e^{-2\pi i/3} Z^{\dagger} \bigr) =
  \begin{pmatrix}
    -g & 0 & 0 \\
    0 & -g & 0 \\
    0 & 0 & 2g
  \end{pmatrix}.
\end{equation}

As shown, the single-excitation sector $\mathcal{H}_{\text{QB},1}$ of the QB ring~\eqref{eq:physical-hamiltonian} indeed realizes the $\Zthree$ Rabi model~\eqref{eq:2-mode-z3-Rabi}. Moreover, the mapping is exact. We note that superconducting qubit platforms are natural for engineering the
interaction~\eqref{eq:physical-hamiltonian}, although analogous constructions are possible for other platforms as well. For example, the optomechanical system implements slightly more general interaction $V_{\text{QB}}$ than that of Eq.~\eqref{eq:physical-hamiltonian}. However, it can still be mapped to the $\Zthree$ Rabi model~\cite{sedov_chiral_2020}. The general $V_{\text{QB}}$ is discussed in App.~\ref{app:arbitrary-interaction-summary}.

\subsubsection*{Why so complicated?}

The suggested implementation of the $\Zthree$ Rabi model from the previous section is considerably more complicated than the setups for the usual $\Ztwo$ Rabi model. Indeed, in order to obtain the $\Ztwo$ Rabi model it is enough to simply couple a qubit with a boson~\cite{forschungszentrumjulichgermany_lecture_2024}.  Therefore, one may wonder if these complications are truly necessary. We argue that they are, because straightforward generalizations of the $\Ztwo$ Rabi model to higher $\Zn$ symmetries encounter fundamental obstacles. In this section, we provide some arguments to explain why this is the case.

There are two most common ways to get a two-level quantum system. It is possible to use a spin as a two-level system  \cite{loss_quantum_1998, bosco_fully_2022, felicetti_quantum_2017, skogvoll_tunable_2021} or two levels in an anharmonic oscillator. For instance, in the latter case the Rabi model describes a superconducting qubit coupled to a cavity~\cite{niemczyk_circuit_2010,forn-diaz_ultrastrong_2017,yoshihara_superconducting_2017,vlasiuk_cavityinduced_2023, kozin_quantum_2024,chen_singlephotondriven_2017,ricco_reshaping_2022}. Therefore, in order to get a $\Ztwo$ Rabi model, one has to take either a spin or an anharmonic oscillator and couple it to a bosonic mode using the interaction term $\hat V = \sigma^x\, \hat x_{\text{boson}}$, where $\sigma^x$ is the Pauli matrix.

Both ways are easily generalized to the $\Zthree$ case. Using a spin-$1$ system might seem natural, but in this case the interaction is realized the spin-$1$ $s^x$ matrix:
\begin{equation}
    s^x= \begin{pmatrix} 0 & \sqrt{2} & 0 \\ \sqrt{2} & 0 & \sqrt{2} \\ 0 & \sqrt{2} & 0 \end{pmatrix}.
\end{equation}
It is easy to see that $s^x$ is not $\Zthree$-symmetric. In other words, the nice properties of the spin-$1/2$ (two-level) Pauli $\sigma^x$ do not generalize to the spin-1 case. This does not prove that building a $\Zthree$ symmetric system (or higher $\Zn$ symmetric systems) out of spins is impossible (we have shown one way to do it in the previous section). However, it illustrates why more intricate constructions are required to realize higher discrete symmetries.

Similarly, if we try to use the three lowest levels of an anharmonic oscillator, the truncated coordinate operator is
not $\Zthree$-symmetric,
\begin{equation}
    \hat x_{\text{an}} = \frac{1}{\sqrt{2}}\begin{pmatrix} 0 & 1 & 0 \\ 1 & 0 & \sqrt{2} \\ 0 & \sqrt{2} & 0 \end{pmatrix}.
\end{equation}
As a result, the $\Ztwo$ Rabi model based on the anharmonic oscillator does not directly generalize to the higher $\Zn$ Rabi models as well.

These obstacles make an experimental realization of $\Zn$ Rabi models non-trivial, which motivates our proposal. The rationale provided in this section is in no way a formal proof. However, the argument should serve as an intuition as to why the $\Zthree$ Rabi model is less straightforward to engineer than its more well-known $\Ztwo$ analog. The model we proposed in the previous section solves this problem by having an inherent $\Zthree$ symmetry originating from the translational symmetry of the qubit ring.

\subsection{Superconducting circuit implementation}
\label{sec:superconducting-implementation}

In this section, we discuss how to implement the QB ring Hamiltonian~\eqref{eq:physical-hamiltonian} in a superconducting (SC) qubit system. The obvious choice for bosonic modes is a cavity (modeled by an LC circuit for our purposes). For the degrees of freedom of the qubits $\sigma_i^{\alpha}$, we consider superconducting charge qubits \cite{bouchiat_quantum_1998,nakamura_coherent_1999,lehnert_measurement_2003,makhlin_quantumstate_2001}, because we require the qubit eigenstates to be charge eigenstates, as we explain in the following. An extensive discussion on the charge qubits that we use is provided in App. \ref{app:charge-qubit}.  Here, we only want to comment that the qubit eigenstates are $|0\rangle$ and $|1\rangle$ defined by $\hat n\, |0\rangle = 0\, |0\rangle, \, \hat n \,|1\rangle = 1\,|1\rangle$, where $\hat n$ is the capacitor charge operator (in units of $2e$). Below, to denote a linear span of these states, we use $\mathcal{V}_k = \operatorname{Span}\{|0\rangle_k, \, |1\rangle_k\}$, where $k$ is the qubit index on the ring.

\begin{figure}[t]
  \includegraphics[width=\linewidth]{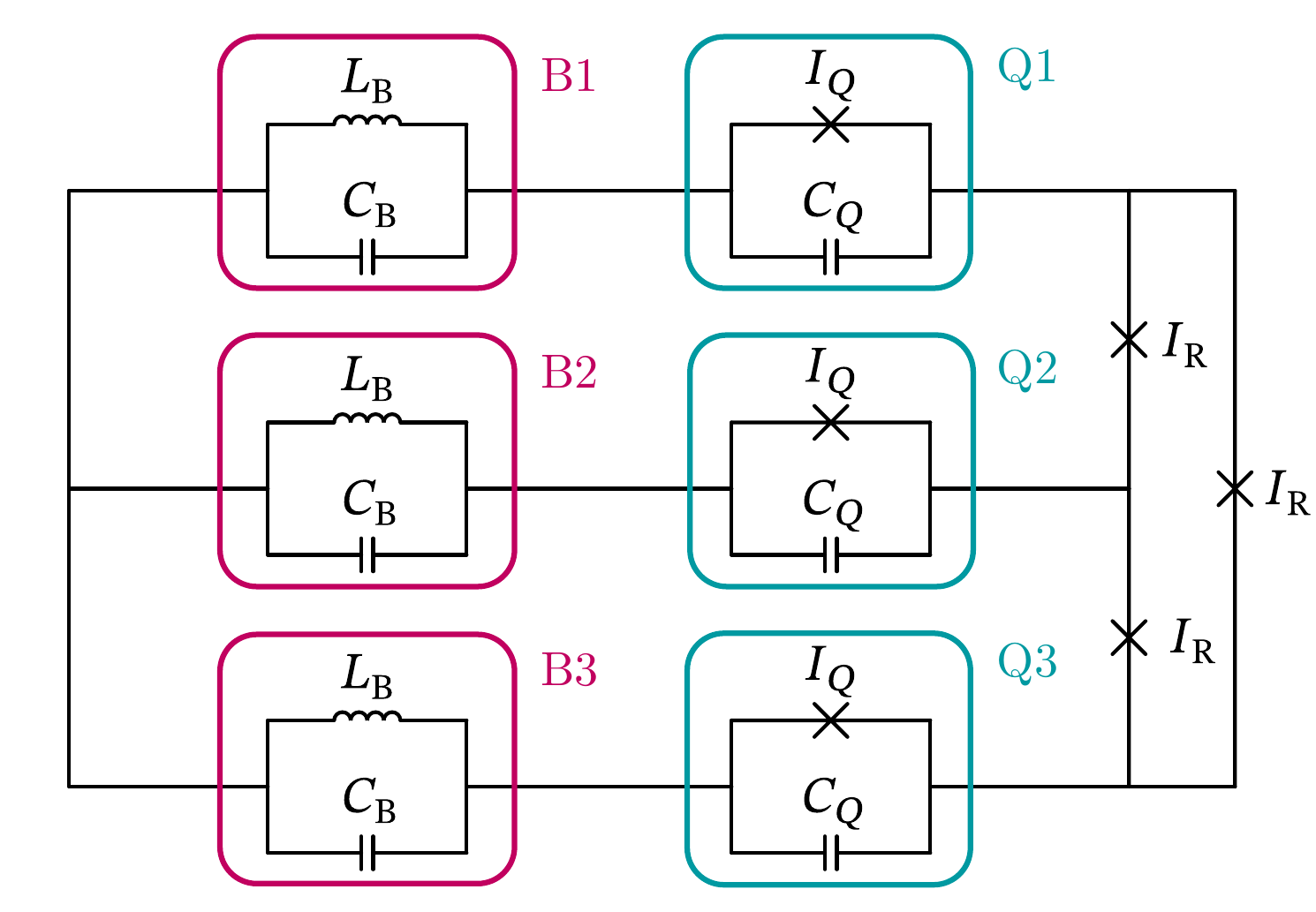}
  \caption{Superconducting circuit implementation of the two-mode $\Zthree$ Rabi model $H_{\text{R}}$ defined in Eq.~\eqref{eq:2-mode-z3-Rabi}. The full circuit is described by the Hamiltonian of the qubit-boson ring $H_{\text{QB}}$, defined by Eq.~(\ref{eq:physical-hamiltonian}), while the $\Zthree$ Rabi model $H_{\text{R}}$ corresponds to its single-excitation subsector. The LC circuits $(L_{\text{B}}, C_{\text{B}})$ host the bosonic modes B1, B2, and B3 (marked by the magenta color); the charge qubits $(I_{\text{Q}}, C_{\text{Q}})$ correspond to the qubit degrees of freedom Q1, Q2, and Q3 (marked by the cyan color); Josephson junctions $I_{\text{R}}$ give rise to the interaction term in the qubit-boson ring Hamiltonian $V_{\text{QB}}$, see Eq.~(\ref{eq:physical-hamiltonian}). Note that the qubit elements here are represented by Cooper pair boxes for simplicity, while the actual implementation requires a more sophisticated variant of a charge qubit, see Appendix~\ref{app:charge-qubit} for details.}
  \label{fig:superconducting-Rabi}
\end{figure}

Figure ~\ref{fig:superconducting-Rabi} shows a superconducting realization of the three‑site QB ring (Sec.~\ref{sec:physical-implementation}). Each horizontal branch of the circuit corresponds to one site $i = 0,1,2$ of the ring. The $i$th boson $\mathrm{B}i$ and qubit $\mathrm{Q}i$ are implemented by the LC circuit and the SC qubit on the $i$th branch, respectively. The Josephson junctions (JJ) in the vertical segments  (right side of the figure) couple the branches and are responsible for the interaction term in $H_{QB}$ from Eq.~\eqref{eq:physical-hamiltonian} \cite{siewert_aspects_2000,rasmussen_controllable_2019,shafranjuk_twoqubit_2006,allman_tunable_2014,hu_controllable_2007}. In Fig.~\ref{fig:superconducting-Rabi}, the characteristic parameters of each circuit element are indicated. Each LC circuit has capacitance $C_{\text{B}}$ and inductance $L_{\text{B}}$. The island of each qubit has a JJ with a critical current $I_{\text{Q}}$ and a capacitance $C_{\text{Q}}$. The JJs in the coupling on the right have a critical current $I_{\text{R}}$. For clarity, we omit any elements that are necessary for an actual experimental implementation, e.g., read‑out resonators, flux‑bias lines, and filtering components.

Next, we argue why this circuit models the Hamiltonian~\eqref{eq:physical-hamiltonian}. Without the coupling JJs, the LC circuits and qubits do not interact with each other and are described by a free Hamiltonian:
\begin{equation}
    H = \sum_{i = 0}^2  H_{\text{LC}, i}(\hat \phi_i, \hat q_i) + \sum_{i = 0}^2 H_{\text{Q},i}(\hat \varphi_i, \hat n_i),
\end{equation}
where $ \hat\phi_i = \hat \Phi_i/\Phi_0$ and $\hat q_i$ are the normalized magnetic flux and  the capacitor charge of the $i$th LC circuit, respectively. Here, $\Phi_0 = h/(2e)$ is the magnetic flux quantum with $h$ being the (non-reduced) Planck constant and $e$ the electron charge. On the other hand,  $\hat \varphi_i$ and $ \hat n_i$ are the JJ superconducting phase and the capacitor charge of the $i$th qubit. The Hamiltonian of the LC circuit is $ H_{\text{LC}, i} = \hat q_i^2/(2C_{\text{B}}) + \hat \phi_i^2/(2L_{\text{B}})$. We leave the microscopic details of the qubit Hamiltonian $H_{Q,i} $ unspecified (beyond being a charge qubit) so that this framework can accommodate different qubit designs. However, we do require that in the two-dimensional subspace spanned by $|0_i\rangle, |1_i\rangle$, the qubit Hamiltonian reduces simply to
\begin{align}
   H_{Q}\bigg|_{\mathcal{V}} = \begin{pmatrix} \bra{0_i} H_Q\ket{0_i} & \bra{0_i} H_Q \ket{1_i} \\ \bra{1_i} H_Q \ket{0_i} & \bra{1_i} H_Q \ket{1_i} \end{pmatrix} = \epsilon \sigma_z,
\end{align}
where the qubit states $\ket{0_i},\ket{1_i}$ are the charge operator eigenstates $ \hat n_i \ket{0_i} = N \ket{0_i}$ and $\hat n_{i} \ket{1}_i = (N+1)\ket{1}_i$ with $ N \in \mathbb{Z}$. The qubit Hilbert space $\mathcal{V}_i = \operatorname{Span}\{\ket{0_i},\, \ket{1_i}\}$ is a computational subspace of the full Hilbert space of the system implementing the qubit, $H_{Q,i}$.  Due to this property, the operator $ e^{i\hat \varphi_i} $ acts on the qubit subspace $\mathcal{V}_i$ as a raising operator:
\begin{equation}
  \begin{aligned}
    &e^{i\hat\varphi_i}\bigg|_{\mathcal{V}_i} = \begin{pmatrix} \bra{0_i} e^{i\hat\varphi_i} \ket{0_i} & \bra{0_i} e^{i\hat\varphi_i} \ket{1_i} \\ \bra{1_i} e^{i\hat\varphi_i} \ket{0_i} & \bra{1_i} e^{i\hat\varphi_i} \ket{1_i} \end{pmatrix} = \sigma^+,
  \end{aligned}
\end{equation}
since we have $[e^{i\hat \varphi}, \hat n] = e^{i\hat \varphi}$. 

Connecting the $i$th and $(i+1)$th circuit branches with a JJ closes a superconducting loop. The flux quantization condition for the loop is
\begin{equation}
  \hat \phi_i + \hat\varphi_i + \hat\phi_R - \hat\varphi_{i+1} - \hat\phi_{i+1} = 2\pi N, \, \text{where}\ N \in \mathbb{Z}.
\end{equation} 
In other words, the superconducting phase $\hat \phi_R$ of the coupling JJ is not an independent degree of freedom; it is fixed by the phases of the adjacent nodes. Therefore, for the corresponding term in the Hamiltonian we obtain:
\begin{equation}
\begin{aligned}
    V_{\text{SC QB},j} &= \frac{\Phi_0 I_{\text{R}}}{2\pi}\cos(\hat \phi_R) \\
    &= \frac{\Phi_0 I_{\text{R}}}{2\pi} \cos(\hat \phi_j + \hat \varphi_j - \hat \phi_{j+1} - \hat \varphi_{j+1}).
\end{aligned}
\end{equation}
Projecting this coupling term onto the two-level subspaces of qubits $j$ and $j+1$ yields the following.
\begin{equation}
\begin{aligned}
    &V_{\text{SC QB},j} \bigg |_{\mathcal{V}_j\otimes \mathcal{V}_{j+1}} = \\
    &=\frac{\Phi_0 I_{\text{R}}}{4\pi}\left(e^{i(\hat\varphi_j - \hat\varphi_{j+1})} e^{i(\hat\phi_j - \hat\phi_{j+1})} + H.c. \right) \bigg|_{\mathcal{V}_j\otimes \mathcal{V}_{j+1}} \\
    &=\frac{\Phi_0 I_{\text{R}}}{4\pi}\left(e^{i(\hat\varphi_j - \hat\varphi_{j+1})} \sigma_j^+ \sigma_{j+1}^- + \mathrm{H.c.} \right).
\end{aligned}
\end{equation}
Here we used the fact that $e^{\pm i\hat\phi}$ act as ladder operators for the charge qubit. This justifies the need specifically for charge qubits rather than flux or phase qubits.

The Josephson junction coupling between two pairs of an LC circuit and a charge qubit gives us precisely the interaction term we wanted. As a result, we conclude that the system depicted in Fig.~\ref{fig:superconducting-Rabi} is described by the Hamiltonian~\eqref{eq:physical-hamiltonian}.  The couplings of the QB ring Hamiltonian can now be expressed via the circuit parameters in the following way:
\begin{equation}
\begin{aligned}
    &\epsilon = \frac{4e^2n_{\mathrm{off}}}{C_Q},\qquad 
    \Omega_{QB} = \frac{1}{\sqrt{L_{\text{B}}C_{\text{B}}}}, \\
    &\eta = \sqrt{\frac{4e^2}{\hbar}}\sqrt[4]{\frac{L_B}{C_B}}, \qquad g = \frac{\Phi_0 I_{\text{R}}}{4\pi}.
\end{aligned}
\end{equation}
For details on the  computation of $\epsilon$, see App.~\ref{app:charge-qubit}. 

Finally, applying the argument from Sec.~\ref{sec:physical-implementation}, we deduce that the circuit is described by the $\Zthree$ Rabi model. Two practical notes are in order. (i) Replacing each coupling JJ by a DC SQUID makes $g \propto I_{\mathrm{R}}$ tunable and allows us to imprint controlled phase offsets around the ring, thereby realizing the ``magnetic'' term with a phase different from that in Eq.~\eqref{eq:superconducting-magnetic-term}. (ii) The device can be initialized at $g\simeq 0$ with the three qubits and three LC modes effectively decoupled; the adiabatic ramp of $g$ then reaches the target operating point, while the standard dispersive readout on the auxiliary resonators provides state discrimination. Taken together, these considerations make the circuit in Fig.~\ref{fig:superconducting-Rabi} a realistic platform for probing $\Zthree$ Rabi model dynamics across coupling regimes without invoking a rotating-wave approximation. The effects of the disorder on the present construction are discussed in App.~\ref{app:disorder}.

\subsection{Optomechanical implementation}
\label{sec:optomechanical-implementation}

\begin{figure}
    \includegraphics[width=\linewidth]{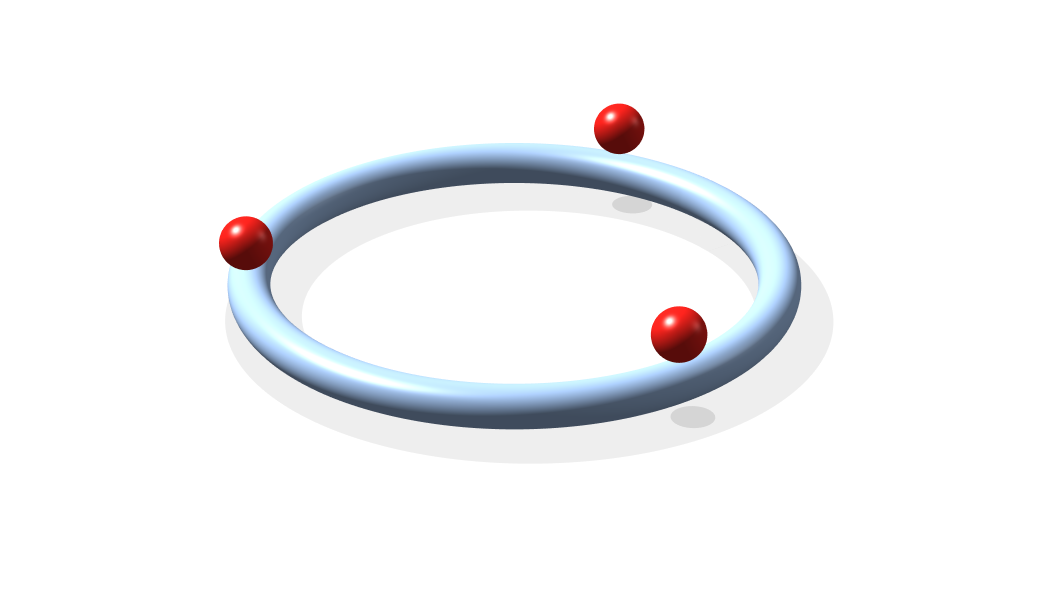}
    \caption{Schematic illustration of the optomechanical implementation~\cite{sedov_chiral_2020} of the two-mode $\Zthree$ Rabi model $ H_{\text{R}'}$, given by Eq.~(\ref{eq:optomechanical-rabi}). The red spheres depict trapped ions carrying two-level systems; their vibrational modes are the bosons from the qubit-boson ring Hamiltonian $H_{\text{QB}}$ in Eq.~(\ref{eq:physical-hamiltonian}). The blue ring is the chiral waveguide that facilitates the interaction in the qubit-boson ring. The full system is described by the Hamiltonian $ H_{\text{OM}}$ as defined in Eq.~\eqref{eq:optomechanical-qb-ring}.}
    \label{fig:optomechanical-rabi}
\end{figure}

Another possible physical platform for the $\Zthree$ Rabi model is an optomechanical system consisting of three ions whose vibration modes are bosonic degrees of freedom. They are connected by a chiral circular waveguide that acts as an interaction medium between the spins and the vibrational phonons (first proposed in~\cite{sedov_chiral_2020}). Although challenging to realize, this platform illustrates the generality of our approach. In this section, we outline how the $\Zthree$ Rabi model arises in this model; for the extended discussion, we refer the reader to the original paper \cite{sedov_chiral_2020}.

The optomechanical model in question is described by the following Hamiltonian:
\begin{equation}
\label{eq:optomechanical-qb-ring}
\begin{aligned}
    H_{\text{OM}}&=\sum\limits_{k} \zeta_k \hat{c}_k^{\dagger} \hat{c}_k + 
 \epsilon \sum_{j=0}^2 \sigma_j^z + \hbar \Omega_{\text{OM}} \sum_{j=0}^2 \hat{b}_j^{\dagger} \hat{b}_j \\
    &+\gamma \sum_{j=0}^2\sum_{k=0}^{\infty} \left[\sigma_j^{+} \hat{c}_k e^{i k\left[R \phi_j+\hat x_j\right]}+\mathrm {H.c.}\right] \text {, }
\end{aligned}   
\end{equation}
where $\hat c_j^\dagger$ ($\hat c_j$) is the chiral waveguide photon creation (annihilation) operator; $\sigma^z_j$, $\sigma_j^+$, and $\sigma_j^-$ act on the ion qubit degree of freedom; $\hat b^{\dagger}_j$ ($\hat b_j$) are the ion vibrational phonon creation (annihilation) operators, with $\hat x_j = \sqrt{\hbar/(2M\Omega_{\text{OM}})}(\hat b_j + \hat b_j^{\dagger})$ being the canonical coordinate operator, and $R\phi_j$ is the equilibrium (average) position of the $j$-th ion, with $R$ being the radius of the waveguide. The waveguide photon dispersion is $\zeta_k = v k$, $\epsilon$ is the Zeeman splitting, $\Omega_{\text{OM}}$ is the phonon frequency, and $\gamma$ is the coupling strength. The photon momentum is quantized and due to the chiral nature of the waveguide, the photons travel only in one direction.

Using the Schrieffer-Wolff transformation \cite{bravyi_schrieffer_2011}, we can integrate out the photon degrees of freedom to get the Hamiltonian in the desired form:
\begin{equation}
\label{eq:optomechanical-transformed-qb-ring}
\begin{aligned}
&H_{\text{OM,eff}}=  \epsilon \sum_{j=0}^2 \sigma_j^z + \hbar \Omega_{\text{OM}} \sum_{j=0}^2 \hat{b}_j^{\dagger} \hat{b}_j \\
& -\frac{\gamma^2R}{2 \hbar v} \sum_{i<j}\left[i \sigma_i^{+} \sigma_j^- e^{i (\epsilon/v) R \phi_{i j}} e^{i \eta\left(\hat{b}_i+\hat{b}_i^{\dagger}-\hat{b}_j-\hat{b}_j^{\dagger}\right)}+\mathrm{H.c.}\right],
\end{aligned}
\end{equation}
where $\eta =(\epsilon/v) \sqrt{\hbar/(2M\Omega_{\text{OM}})}$, $M$ is the mass of the trapped atoms, and $\phi_{ij}=\phi_i-\phi_j$.
The Hamiltonian is a bit different from Eq.~\eqref{eq:physical-hamiltonian}. It is treated in App. \ref{app:arbitrary-interaction-summary} and gives rise to the following $\Zthree$ Rabi model:
\begin{equation}
\label{eq:optomechanical-rabi}
\begin{aligned}
    &H_{\text{R}'}= - \frac{\gamma^2R}{2 \hbar v} (e^{-\pi i/6} Z + e^{\pi i/6} Z^{\dagger}) \\
    &+ \hbar \Omega_{\text{OM}} (\hat a_1^{\dagger} \hat a_1 + \hat a_2^{\dagger} \hat a_2) \\
    &- \frac{\hbar \Omega_{\text{OM}} \eta}{6^{1/2}}\left[(\hat a_1 + \hat a_2^{\dagger}) X  + (\hat a_2 + \hat a_1^{\dagger}) X^{\dagger}  \right]
\end{aligned}
\end{equation}
which differs from the $\Zthree $ Rabi model we obtained in the superconducting circuit realization only by the finite phase $\varphi = - 5\pi/6$ of the $\Zthree$ ``magnetic'' term.
For convenience we provide the explicit matrix realization of the ``magnetic'' term: 
\begin{equation}
\label{eq:optomechanical-magnetic-term}
    \frac{-\gamma^2 R}{2\hbar v} (e^{-\pi i/6} Z + e^{\pi i/6} Z^2) = \frac{\sqrt{3}\gamma^2 R}{2\hbar v}\begin{pmatrix}
         1& 0 & 0 \\ 0 & -1 & 0 \\ 0 & 0 & 0 
    \end{pmatrix}.
\end{equation}
As one can see, the non-zero phase $\varphi$ leads to all the eigenvalues being non-degenerate [cf. Eq.~\eqref{eq:superconducting-magnetic-term}].

As a result, the optomechanical system in Eq.~\eqref{eq:optomechanical-qb-ring} provides another way to realize the $\Zthree$ Rabi model~\eqref{eq:2-mode-z3-Rabi}. This implementation uses the chiral waveguide as a mediator for the qubit-boson interaction $V_{\text{QB}}$~\eqref{eq:physical-hamiltonian}, which is a bit more complicated compared to the superconducting implementation (Fig.~\ref{fig:superconducting-Rabi}). Nevertheless, the optomechanical system~\eqref{eq:optomechanical-qb-ring} still realizes the Rabi model through the mechanism described in Sec.~\ref{sec:physical-implementation}.

\section{\texorpdfstring{$\Zthree$}{Z3} Potts model}
\label{sec:potts-model}

\subsection{Theoretical description}
\label{sec:theoretical-potts}

The $\Zthree$ Potts model \cite{wu_potts_1982} is a one-dimensional chain of three-level systems (qutrits) with a global $\Zthree$ symmetry. Originally, it appeared in the context of statistical physics as a generalization of the Ising model  \cite{wu_potts_1982,baxter_critical_1982}. Today, the Potts model is relevant for qudit-based quantum computation~\cite{aharonov_polynomial_2007, okada_efficient_2019}. However, direct experimental realization of the Potts model has not yet been achieved. One of the goals of this paper is to propose an experimental realization using $\Zthree$ Rabi models as building blocks. To achieve this, we generalize the idea of building an Ising model by coupling a chain of $\Ztwo$ Rabi models, proposed by Hwang \cite{hwang_largescale_2013}, to the $\Zthree$-symmetric case.

The Hamiltonian of the $\Zthree$ Potts model with open b.c. is given by 
\begin{equation}
\begin{aligned}
\label{eq:potts-hamiltonian}
H_{\text{P}} =& f_{\text{P}} \sum\limits_{m=1}^L \left(e^{i\phi}Z_m + e^{-i\phi}Z_m^{\dagger}\right) \\
  &+  J_{\text{P}} \sum\limits_{m=1}^{L-1} \left(X_m X_{m+1}^{\dagger} + X_{m+1} X_{m}^{\dagger}\right),
\end{aligned}
\end{equation}
where $X_m$ and $Z_m$ are the shift and clock matrices~\eqref{eq:shift-clock-matreces} that act on the $m$th qutrit. Here, $f _\text{P}$ describes the strength of a single-site potential, $J_{\text{P}}$ is the coupling strength of $\Zthree$-symmetric nearest-neighbor interaction.

As we have already discussed, it is difficult to obtain the symmetry $\Zthree$ in a chain of an arbitrary three-level system. However, we have already learned how to construct a $\Zthree$ Rabi model, which we can now use as a building block for the $\Zthree$ Potts model. 
In the rest of this section, we proceed in the following way.
We first assemble a chain of the $\Zthree$ Rabi models. 
Then, the three cat-states in the  $m$th $\Zthree$ Rabi model~\eqref{eq:three-cat-states} are identified with the three states on the $m$th site of the Potts model $ |j\rangle_m = |\psi_j\rangle_m $ for $j=0,\, 1,\, 2$.
In the extreme-coupling limit of the Rabi model~\eqref{eq:2-mode-z3-Rabi}, $ \lambda \gg \hbar \Omega_{\text{R}}$,  the higher energy states are effectively decoupled from the $|\psi_j\rangle$ states.

It is convenient to build the Potts model from the Rabi models, since the boson degrees of freedom in the latter allow us to obtain a $\Zthree$-symmetric interaction. 
The reasoning is simple: the Rabi model's $\Zthree$ symmetry acting on bosons is a residue of a usual boson $\mathrm{U(1)}$ symmetry. 
Consequently, a $\mathrm{U(1)}$-symmetric coupling between the Rabi models also satisfies the $\Zthree$ symmetry. The simplest $\mathrm{U(1)}$-symmetric boson interaction is just a hopping term $\hat a_m^{\dagger} \hat a_{m+1} + \textrm{H.c.}$ As a result, in order to obtain the $\Zthree$ Potts model we need to take a chain of $\Zthree$ Rabi models and couple them by the boson hopping term:
\begin{equation}
\label{eq:coupled-rabi}
\begin{aligned}
    &H_{\text{R chain}} = \sum\limits_{m=1}^L H_{\text{R}, m} \\
    &+ J \sum\limits_{m=1}^{L-1} \sum\limits_{k=1}^2\left( \hat a_{m,k}^{\dagger} \hat a_{m+1,k} + \hat a_{m+1,k}^{\dagger} \hat a_{m,k}\right),
\end{aligned}
\end{equation}
where $m =1, \dots, L$ labels the chain sites and $k=1,2$ labels  We claim that this Hamiltonian gives us precisely the $\Zthree$ Potts model when we restrict it to the Hilbert space $\bigotimes_m\mathcal{R}_m$ generated by the three cat states on the $m$th site.

Within the Rabi qutrit subspace $\mathcal{R}_m$, the operator $\hat a_{m,k}$ acts as a permutation between the cat states, as can be seen from Eq.~\eqref{eq:three-cat-states}: 
\begin{equation}
\begin{aligned}
    \hat a_{m,1} |\psi_i\rangle_m &= (\lambda/\hbar \Omega_{\textrm{R}})|\psi_{i+1}\rangle_m,\\
    \hat a_{m,2} |\psi_i\rangle_m &= (\lambda/\hbar \Omega_{\textrm{R}})|\psi_{i-1}\rangle_m.
\end{aligned}\end{equation}
The $1$st and $2$nd mode annihilation operators act as opposite cyclic permutations. The action of the creation operator is slightly more complicated:
\begin{equation}
\begin{aligned}
    &\hat a_{m,1}^{\dagger} |\psi_i\rangle_m =  (\lambda/\hbar \Omega_{\textrm{R}})|\psi_{i-1}\rangle_m + \ket{\delta_{\perp,1}},\\
    &\hat a_{m,2}^{\dagger} |\psi_i\rangle_m =  (\lambda/\hbar \Omega_{\textrm{R}})|\psi_{i+1}\rangle_m + \ket{\delta_{\perp,2}},
\end{aligned}
\end{equation} 
where $\ket{\delta_{\perp,1}}$ and $\ket{\delta_{\perp,2}}$ are components orthogonal to the cat-state subspace $\mathcal{R}_m$, which satisfy $\bra{\delta_{\perp,k}}\ket{\delta_{\perp,k}} = 1$. 
In the extreme-coupling regime $\lambda/\hbar \Omega_{\text{R}} \gg 1$ (see Refs.~\cite{kozin_cavityenhanced_2025,KozinMiserevSchottky}), we can neglect $\ket{\delta_{\perp}}$. As a result, within the Rabi qutrit subspace $\mathcal{R}_m$, both the creation and annihilation operators act as cyclic permutations, 
\begin{equation}
\begin{aligned}
    \hat a_{m,1}|_{\mathcal{R}_m} &= (\lambda/\hbar \Omega_{\textrm{R}}) X_m, \\
    \hat a_{m,2}|_{\mathcal{R}_m} &= (\lambda/\hbar \Omega_{\textrm{R}}) X^{\dagger}_m, \\
    \hat a_{m,1}^{\dagger}|_{\mathcal{R}_m} &\approx (\lambda/\hbar \Omega_{\textrm{R}}) X^{\dagger}_m, \\
    \hat a_{m,2}^{\dagger}|_{\mathcal{R}_m} &\approx (\lambda/\hbar \Omega_{\textrm{R}}) X_m.
\end{aligned}
\end{equation}
 Consequently, in this sector the chain of coupled Rabi models is equivalent to the Potts model:
\begin{equation} \label{eq:effective-Potts-coupling}
\begin{aligned}
    J &\sum\limits_{k=1}^{2} \left(\hat a_{m,k}^{\dagger} \hat a_{m+1,k} + \hat a_{m+1,k}^{\dagger} \hat a_{m,k}\right) = \\
    &2(\lambda/\hbar \Omega_{\textrm{R}})^2 J\left(X_m X_{m+1}^{\dagger} + X_{m+1} X_m^{\dagger}\right).
\end{aligned}
\end{equation} 
In summary, a chain of $\Zthree$ Rabi models coupled by boson hopping implements the nearest-neighbor term of the $\Zthree$ Potts Hamiltonian, in which the degrees of freedom of the qutrit are formed by the $\Zthree$ Rabi cat states~\eqref{eq:three-cat-states}. Comparing Eqs.~(\ref{eq:effective-Potts-coupling}) and (\ref{eq:potts-hamiltonian}), we see that the coupling parameters are related as $J_{\text{P}} = 2(\lambda/\hbar \Omega_{\textrm{R}})^2 J$. 
On the other hand, the single-site terms $f_{\text P}(e^{i\phi}Z_m+\mathrm{H.c.})$ originate from the small energy splitting of the $\Zthree$ cat states [Eq.~\eqref{eq:rabi-spectrum}], which is controlled by the parameter $B$. The mapping requires the extreme-coupling regime $\lambda/\hbar \Omega_{\mathrm R}\gg 1$ and the scale separation $J_{\text P},f_{\text P}\ll \hbar \Omega_{\text{R}}$. 

\subsubsection*{Qubit-boson-ring point of view}
\label{sec:underlying-qb-ring}

 Previously, we showed how to construct the $\Zthree$ Potts model out of a chain of $\Zthree$ Rabi models. However, each Rabi model is implemented by a QB ring [see Sec.~\ref{sec:physical-implementation}]. Here, we show how these QB rings should be coupled to obtain the boson-hopping~\eqref{eq:coupled-rabi}. It turns out to be quite straightforward. The hopping terms between the QB ring bosons $\hat b_{m,j}$ and $\hat b_{m,j}^{\dagger}$ give us precisely the hopping terms between the $\Zthree$ Rabi model bosons $\hat a_{m,k}$ and $\hat a_{m,k}^{\dagger}$ since the Fourier transform does not change the form of the boson hopping:
\begin{equation}
\begin{aligned}
    &\sum\limits_{j=0}^2 \left(\hat b_{m,j}^{\dagger} \hat b_{m+1,j} + \hat b_{m+1,j}^{\dagger} \hat b_{m,j}\right) \\
    =&\sum\limits_{k=0}^{2} \left(\hat b_{m}^{\dagger}(k) \hat b_{m+1}(k) + \hat b_{m+1}^{\dagger}(k) \hat b_{m}(k)\right),
\end{aligned}
\end{equation} 
where $\hat b_{m}(k)$ and $\hat b_m^{\dagger}(k)$ are the Fourier transformed QB ring bosonic modes. 
Taking into account that {\it by definition} we have $\hat a_{m,k} = \hat b_m(k)$ and $\hat a_{m,k}^{\dagger} = \hat b^{\dagger}_m(k)$  for $k =1$ and $2$, as written after Eq.~\eqref{eq:effective-rabi},  we immediately obtain the desired form of the interaction~\eqref{eq:coupled-rabi}. Also, we have a hopping term for the zero-momentum bosonic mode $\hat b_{m}(0)$ and $\hat b^{\dagger}_{m}(0)$. Since it decouples from  the rest of the system (cf. Sec. \ref{sec:physical-implementation}), we ignore it. 

Using Eqs.~(\ref{eq:rabi-spectrum}), (\ref{eq:QB-RM-parameter-mapping}), and (\ref{eq:effective-Potts-coupling}), one can check that the couplings in the Potts Hamiltonian can be related to the underlying QB ring parameters in the following way:
\begin{equation}
\begin{aligned}
    &f_{\text{P}} = g \exp\left(-\frac{\eta^2}{2}\right),\\
    &J_{\text{P}} = \frac{\eta^2 J }{3}.
\end{aligned}
\end{equation}

Thus, we see that the $\Zthree$ Potts model can be realized using a chain of the QB ring units, each implementing the $\mathbb{Z}_3$ Rabi model. In the following, we illustrate this proposal with a specific superconducting circuit (Fig.\ref{fig:superconducting-potts}).

\subsection{Coupled superconducting \texorpdfstring{$\Zthree$}{Z3} Rabi models as the Potts model}

\begin{figure}[t]
    \includegraphics[width=\linewidth]{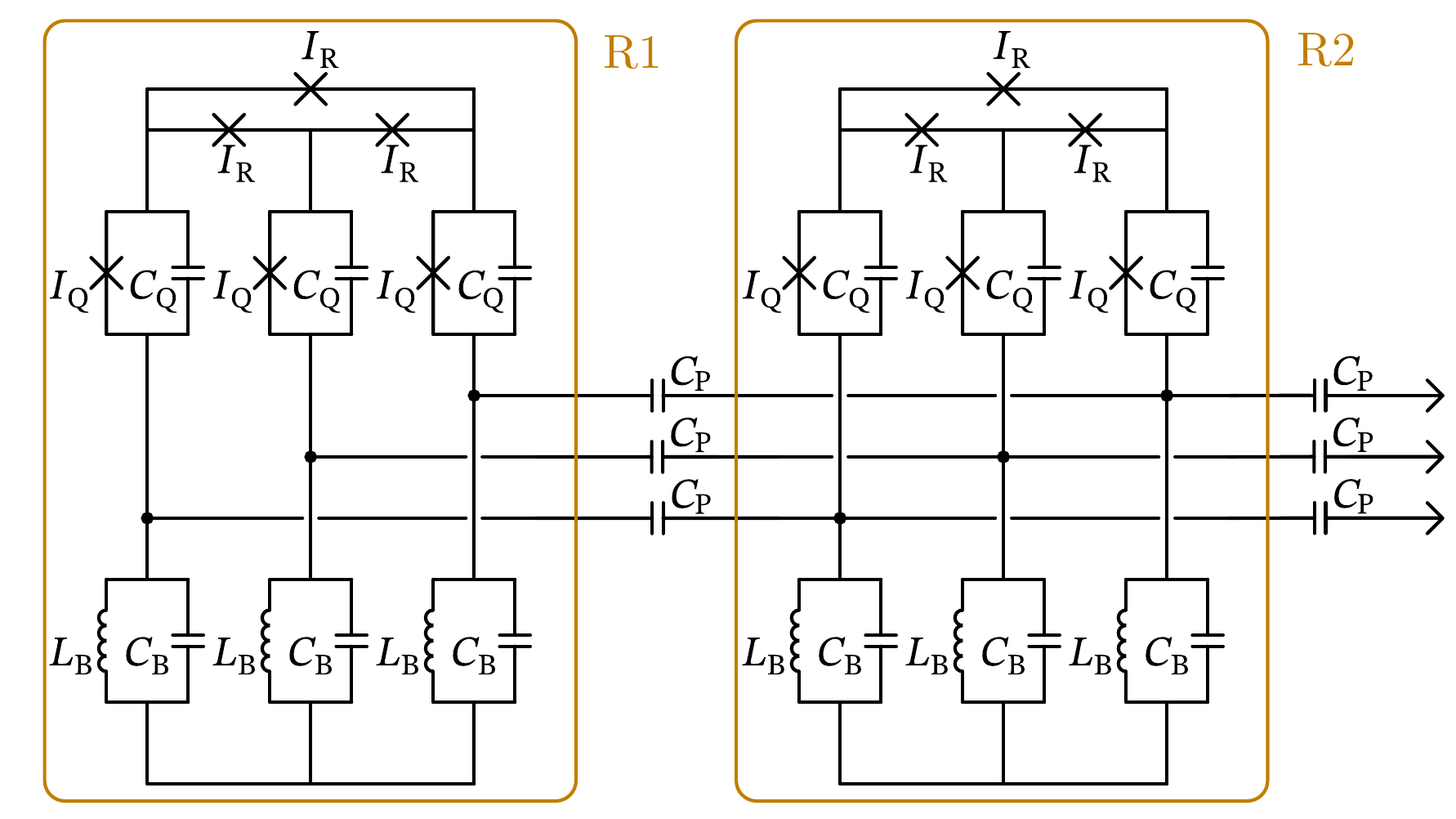}
    \caption{Superconducting circuit implementation of the $\Zthree$ Potts model (\ref{eq:potts-hamiltonian}). It is composed from a chain of coupled superconducting circuits $\mathrm{Ri}$, with $i = 1,2, \ldots, L$ (only the first two are shown explicitly in the figure), each representing the $\Zthree$ Rabi model [see Fig.~\ref{fig:superconducting-Rabi}].  
    The full circuit is described by the Hamiltonian~\eqref{eq:coupled-rabi}. In the notation for circuit elements, the subscript ``B'' denotes elements that correspond to boson degrees of freedom; the subscript ``Q'' denotes elements hosting qubits; the subscript ``R'' denotes elements responsible for the interaction in the qubit-boson ring; finally, the subscript ``P'' denotes elements that give rise to the interaction in the Potts model. }
    \label{fig:superconducting-potts}
\end{figure}

Having explored the theoretical construction of the Potts model based on a chain of coupled Rabi models, here we show how to implement the Potts model using superconducting circuits. We consider a chain of superconducting circuits that simulate the $\Zthree$ Rabi models [see Sec.~\ref{sec:superconducting-implementation}]. Next, we need to couple the neighboring Rabi models with a boson hopping term to get the Hamiltonian~\eqref{eq:coupled-rabi}. As we have shown in Sec.~\ref{sec:underlying-qb-ring}, we must couple the boson modes of the neighboring QB rings pairwise to achieve this. Therefore, we connect the corresponding LC circuits of the neighboring QB rings by capacitive coupling (Fig.~\ref{fig:superconducting-potts}). 

After going from the Lagrangian to the Hamiltonian description, the nearest-neighbor capacitive coupling induces an all-to-all coupling among the canonical momenta. However, this complication can be ignored assuming that the capacitance responsible for the coupling of the nearest-neighbor Rabi models is much smaller than the cavity capacitance, $C_{\text{P}} \ll C_{\text{B}}$~\cite{forschungszentrumjulichgermany_lecture_2024}. 
In this limit, the nearest-neighbor coupling turns out to be proportional to $C_{\text{P}}/C_{\text{B}}^2$, capacitors are: 
\begin{align} \label{eq:capacitive-coupling}
    &V_{\text{P},m,j} =
    -\frac{C_{\text{P}}}{2C_{\text{B}}^2} (\hat q_{m+1,j} - \hat q_{m,j})^2 \\
    &= -\frac{2e^2C_{\text{P}}}{C_{\text{B}}^2}\left(\hat n_{m,j}^2 + \hat n_{m+1,j}^2\right)\nonumber\\
    &- \frac{e^2C_{\text{P}}}{\eta^2C_{\text{B}}^2}\left(\hat b^{\dagger}_{m,j} \hat b^{\dagger}_{m+1,j} - \hat b^{\dagger}_{m,j} \hat b_{m+1,j} + \mathrm{H.c.}\right),\nonumber
\end{align}
where we expressed the charge $\hat q_{m,j}$ through the creation and annihilation operators, $\hat q_{m,j} = 2e\hat n_{m,j} = i2e \eta(\hat b_{m,j}^{\dagger} - \hat b_{m,j})/\sqrt{2}$.

We now neglect the terms $ \hat n_{m,j}^2 $ and $ \hat n_{m+1, j}^2 $ in Eq.~(\ref{eq:capacitive-coupling}) because they simply renormalize the parameters of the Rabi model. Additionally, the capacitor coupling produces unbalanced terms $\hat b^{\dagger}_{m,j} \hat b^{\dagger}_{m+1,j}$ and $\hat b_{m,j} \hat b_{m+1,j}$. However, we can eliminate them with the rotating wave approximation (RWA). For RWA to be applicable, we require $\hbar\Omega_{QB} \gg e^2 C_{\text{P}}/\eta^2 C_{\text{B}}^2$, which, after simplification, turns out to be equivalent to $C_{\text{P}} \ll C_{\text{B}}$. After applying the RWA, we are left precisely with the hopping term between the bosonic modes of the neighboring Rabi models. Consequently, according to Sec.~\ref{sec:theoretical-potts}, the superconducting circuit in Fig.~\ref{fig:superconducting-potts} implements the $\Zthree$ Potts model. The coupling parameters of the resulting Potts model are related to the circuit parameters as follows.
\begin{equation}
\begin{aligned}
    &f_{\text{P}} = \frac{\Phi_0I_{\text{R}}}{4\pi} \exp\left(-\frac{2e^2}{\hbar}\sqrt{\frac{L_{\text{B}}}{C_{\text{B}}}}\right), \\
    &J_{\text{P}} = \frac{e^2 C_{\text{P}}}{3C_{\text{B}}^2}. 
\end{aligned}
\end{equation}

In summary, we propose the superconducting circuit (see Fig.~\ref{fig:superconducting-potts}) that simulates the $\Zthree$ Potts model. The required parameter range includes: 
\begin{enumerate}[label=(\roman*)]
    \item Rabi extreme-coupling regime, 
    \begin{equation}
    \tag{\theequation .A}
    \frac{\eta}{\sqrt{6}} = \sqrt{\frac{2e^2}{3\hbar}} \sqrt[4]{\frac{L_{\text{B}}}{C_{\text{B}}}}\gg 1;
    \end{equation}
    \item Potts model parameters must keep the dynamics within the cat-state qutrit subspace, 
    \begin{equation}
    \tag{\theequation .B}
    \begin{aligned}J_{\text{P}}=\frac{e^2 C_{\text{P}}}{3C_{\text{B}}^2}&\ll \frac{\hbar}{\sqrt{L_{\text{B}}C_{\text{B}}}}, \\
    f_{\text{P}} = \frac{\Phi_0I_{\text{R}}}{4\pi} \exp\left(-\frac{2e^2}{\hbar}\sqrt{\frac{L_{\text{B}}}{C_{\text{B}}}}\right) &\ll \frac{\hbar}{\sqrt{L_{\text{B}}C_{\text{B}}}};
    \end{aligned}
    \end{equation}
    \item RWA being valid and the capacitive coupling being local, 
    \begin{equation}
    \tag{\theequation .C}
    \frac{C_{\text{P}}}{C_{\text{B}}} \ll 1.
    \end{equation}
\end{enumerate} 
All these requirements should be achievable for the present stage of superconducting-circuit platforms. As a result, we believe that our setup is feasible for experimental realization.

\subsection{Coupled optomechanical \texorpdfstring{$\Zthree$}{Z3} Rabi models as the Potts model}

\begin{figure}[t]
    \centering
    \includegraphics[width =  \linewidth]{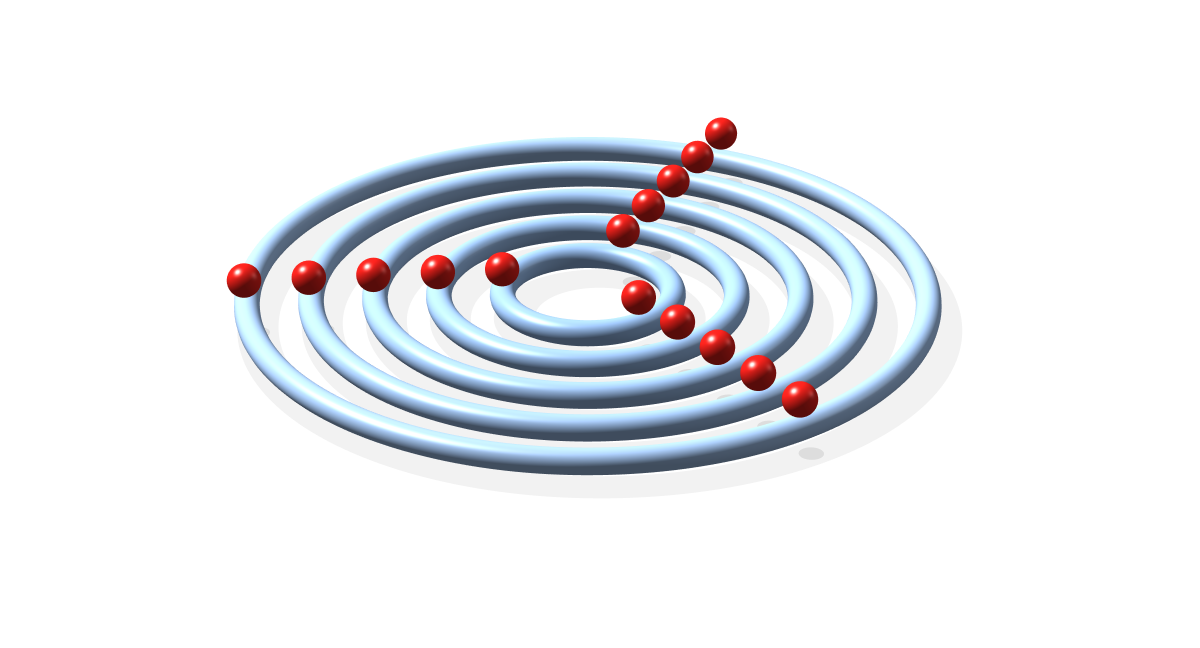}
    \caption{Schematic illustration of the optomechanical (OM) implementation of the Potts model. It consists of several concentric OM implementations of the Rabi model, i.e., several concentric cyclic chiral waveguide and three trapped ions on top of each waveguide. The notations follow those in Fig.~\ref{fig:optomechanical-rabi}. }
    \label{fig:optomechanical-potts}
\end{figure}

In principle, we can construct an optomechanical implementation of the $\Zthree$ Potts model by similarly coupling multiple $\Zthree$ Rabi units. However, the geometry is a bit more peculiar due to the use of a circular waveguide in the optomechanical Rabi setup. To build the optomechanical $\Zthree$ Potts model we arrange the optomechanical $\Zthree$ Rabi models concentrically (see Fig. \ref{fig:optomechanical-potts}). On the one hand, the radius of the waveguide does not affect the parameters of the model. Furthermore, placing the ions of neighboring Rabi models close to each other generates an effective phonon-phonon coupling via the Coulomb interaction \cite{schneider_experimental_2012,timm_dynamics_2023}. The phonon-phonon coupling between the neighboring Rabi models leads to the same effective Hamiltonian~\eqref{eq:coupled-rabi}, realizing the $\Zthree$ Potts model as described in Sec.~\ref{sec:potts-model}.

\subsection{Chiral clock model and parafermions}

In this subsection, we briefly review the $\Zthree$ chiral clock model. It is a generalization of the $\Zthree$ Potts model that has a chiral nearest-neighbor interaction:
\begin{equation}
\label{eq:chiral-potts-hamiltonian}
\begin{aligned}
    H_{\text{cl}} = &f'\sum_{m=1}^L (e^{i\phi'} Z_m + e^{-i\phi'} Z_m^{\dagger}) \\
    &+ J' \sum_{m=1}^{L-1} (e^{i\theta'} X_m X_{m+1}^{\dagger} + e^{-i\theta'} X_{m+1}X_{m}^{\dagger} ).
\end{aligned}
\end{equation} 
In other words, it is obtained from the $\Zthree$ Potts model~\eqref{eq:potts-hamiltonian} by allowing for a complex nearest-neighbor coupling $J'e^{i\theta'}$. 

In the construction described in Sec.~\ref{sec:potts-model}, the interaction parameter $J'$ originates from the interaction between bosons of the neighboring QB rings. Hence, if we want to engineer the chiral clock model, we have to find a way to create a chiral boson hopping. Chirality arises when hopping between bosonic modes acquires a complex amplitude, so that excitations accumulate a nontrivial phase when moving around a closed loop. In general, such a complex hopping term
\begin{equation}
H_{nm} = g \, e^{i\phi_{nm}} \hat a_n^\dagger \hat a_m + \mathrm{H.c.}
\end{equation}
corresponds to a synthetic gauge field, with the directed phase factors $\phi_{nm}$ acting as Peierls phases. The net phase around a plaquette defines an effective magnetic flux that breaks time-reversal symmetry and leads to the chiral transport of photons or phonons.

In trapped-ion arrays, the bosons are vibrational phonons. Coupling between the phonons is engineered through the Coulomb interaction combined with laser-induced forces. By introducing a static gradient in trap frequencies and applying periodic modulation (photon-assisted tunneling), the effective phonon hopping becomes resonant and inherits the phase of the driving field. Adjusting laser phases thus imprints controlled Peierls phases on the phonon couplings, producing synthetic fluxes that enable chiral phonon propagation \cite{bermudez_synthetic_2011, schneider_experimental_2012,vermersch_implementation_2016,roushan_chiral_2017,kiefer_floquetengineered_2019,bazavan_synthetic_2024}.

In superconducting circuits, the bosons are microwave photons in resonators or qubits. Here, Josephson junctions provide nonlinear couplers whose effective interaction strength can be modulated parametrically by flux or multi-tone microwave pumping. When the pump frequency matches the detuning between two modes, frequency-converting tunneling occurs with an amplitude set by the pump strength and a phase set by the pump phase. Embedding such complex hoppings in a ring of modes yields a net synthetic magnetic flux, leading to unidirectional photon circulation \cite{kapit_quantum_2013,sliwa_reconfigurable_2015,gu_microwave_2017, cao_parametrically_2024}.

Assuming that we obtained the chiral coupling in one of the ways described above, we now want to discuss what physical effects the complex coupling causes in the chiral clock model. The most interesting consequence is parafermion physics~\cite{fendley_parafermionic_2012, ronetti_clock_2021, calzona_$mathbbz_4$_2018a,mazza_nontopological_2018}. The duality between the chiral clock model and the parafermion chain is a generalization of the duality between the Ising model and the Kitaev chain that displays Majoranas \cite{kitaev_unpaired_2001}. 
The mapping between the degrees of freedom of qutrit and parafermion is achieved via a generalization of the Jordan-Wigner transformation, the so-called Fradkin-Kadanoff (FK) transformation \cite{fradkin_disorder_1980}.
\begin{equation}
\Gamma_m = \prod\limits_{m'<m} Z_{m'} X_m , \quad \Delta_m = \prod\limits_{m'\le m} Z_{m'} X_m.
\end{equation}
Here, $\Gamma_j, \, \Delta_j$ are parafermion operators that satisfy the following commutation relations: $\Gamma_m \Gamma_{m'} = \omega^{\sgn(m-m')} \Gamma_{m'} \Gamma_{m}, \, \Delta_m \Delta_{m'} = \omega^{\sgn(m-m')} \Delta_{m'} \Delta_m$, and $\Gamma_m \Delta_{m'} = \omega^{\sgn(m-m')} \Delta_{m'} \Gamma_m$.

Using the FK transformation, it is easy to show that the chiral clock model is equivalent to a parafermion chain. The issue is that the FK transformation is non-local. In terms of parafermion operators, the chiral clock model Hamiltonian~\eqref{eq:chiral-potts-hamiltonian} reads
\begin{equation}
\begin{aligned}
    H_{\text{chP}} = &f' \sum_{m=1}^L\left( e^{i\phi'} \Gamma_m \Delta_m^{\dagger} + e^{-i\phi'} \Delta_m \Gamma_m^{\dagger}\right) \\
    &+J' \sum_{m=1}^{L-1} \left(e^{i\theta'} \Delta_m \Gamma_{m+1}^{\dagger} + e^{-i\theta'} \Gamma_{m+1} \Delta_m^{\dagger}\right).
\end{aligned}
\end{equation}

On the other hand, it is well-known that the parafermion chain hosts a topological phase with the parafermion edge modes  \cite{fendley_parafermionic_2012}:
\begin{equation}
\label{eq:parafermion-edge-mode}
\Gamma_{\text{edge}} = \Gamma_1 + \frac{f'}{J' \sin(3\theta')} \mathcal{O} + \dots,
\end{equation}
where $\mathcal{O}$ is a known combination of parafermion operators at the first lattice site, and the ellipses represent higher order terms.
This expression also clarifies why we require the chiral coupling~\eqref{eq:chiral-potts-hamiltonian}. For the parafermion edge mode to be localized, the expansion parameter must satisfy $f'/(J'\sin(3\theta')) < 1$. However, in the non-chiral case, $\theta' = 0 $, the expansion parameter diverges.

\section{Conclusion}
\label{sec:conclusion}
We established a hierarchy of $\Zthree$-symmetric models. First, we have shown that the single-excitation sector of the qubit–boson ring (\ref{eq:physical-hamiltonian}) can be mapped to the $\Zthree$ Rabi model~(\ref{eq:2-mode-z3-Rabi}). Then, we considered a chain of coupled $\Zthree$ Rabi models, each hosting an individual qutrit degree of freedom. The resulting chain simulates the $\Zthree$ Potts model. In the QB ring, the $\Zthree$ symmetry appears as a discrete rotation; upon projecting to the single-excitation manifold, it acts as an internal symmetry that is inherited by the coupled-block construction of the $\Zthree$ Potts Hamiltonian.

Our main goal was to suggest a way towards an experimental realization of the $\Zthree$ Rabi and Potts models. The superconducting circuit we proposed is experimentally feasible. However, this approach does not have to be limited only to superconducting-circuit platforms. For instance, we believe that a similar strategy could work for spin-qubit systems, though this remains a topic for future research.

More broadly, our work highlights the intriguing potential of $\Zn$-symmetric systems within condensed matter physics. These systems present fertile ground for discovering novel quantum phenomena, many of which remain to be fully understood or described.

\section*{Acknowledgments} 

The authors thank Daria Kalacheva, Henry Legg, Katharina Laubscher, and Ilia Luchnikov for fruitful discussions and useful comments. This work was supported as part of NCCR SPIN, a National Center of Competence in Research, funded by the Swiss National Science Foundation (grant number 225153). This work has received funding from the Swiss State Secretariat for Education, Research and Innovation (SERI) under contract number M822.00078. D.L. acknowledges the Deanship of Research and the Quantum Center  for the support received under Grant no. CUP25102 and no. INQC2500, respectively.

\appendix

\section{QB ring generalization to an arbitrary interaction matrix}
\label{app:arbitrary-interaction-summary}

In this appendix, we summarize how the derivation in Sec.~\ref{sec:physical-implementation} extends to a QB ring
with the most general $\Zthree$-symmetric form of the interaction between different qubit-boson pairs. The system is described by a direct generalization of the Hamiltonian~\eqref{eq:physical-hamiltonian}, which reads  
\begin{equation}
\label{eq:arbitary-interaction-hamiltonian}
  \begin{aligned}
    H_{\text{gen}} &= \epsilon \sum_{j=0}^{2} \sigma_j^z
      + \hbar \Omega \sum_{j=0}^{2} \hat a_j^{\dagger} \hat a_j + V_{\text{gen}},
      \\
    V_{\text{gen}} &= g \sum_{j,k=0}^2 A_{jk}
      \sigma_j^{+} \sigma_k^{-} e^{ i ( \hat x_j - \hat x_k ) },
  \end{aligned}
\end{equation}
where the position indices are periodic, $j = k = 0 =3$, and the interaction matrix is Hermitian $A^{\dagger}=A$. Applying the same sequence of steps as in Sec.~\ref{sec:physical-implementation}, i.e., the momentum translation $S$,
the Fourier transform, the restriction to the single-excitation sector $\mathcal H_{\text{QB},1}$, and final basis transformations, we obtain a Hamiltonian that has the same form as that in Eq.~\eqref{eq:effective-rabi}. 
The only difference is that the qutrit term $g(e^{2\pi i/3}Z+e^{-2\pi i/3}Z^{\dagger})$ in Eq.~(\ref{eq:effective-rabi}) is replaced by $gX^{\dagger}HAH^{\dagger}X$, where $H$ is the Hadamard matrix.
Taking into account that $A$ is Hermitian and commutes with $\Zthree$ symmetry, the qutrit term can be written in the canonical form $gHAH^{\dagger} = \tilde g(e^{i\phi} Z + e^{-i\phi}Z^{\dagger})$, where the parameters $\tilde g$ and $\phi$ depend on the matrix $A$.

In the superconducting implementation, one has  $A = X + X^{\dagger}$ (here $A$ is a $3\times 3$ matrix), thereby
recovering the special case treated in Sec.~\ref{sec:physical-implementation}. On the other hand, the optomechanical realization of the QB ring in Eq.~\eqref{eq:optomechanical-transformed-qb-ring}
gives rise to a different interaction matrix, $A = -iX + iX^{\dagger}$. Nevertheless, the logic stays exactly the same. As a result, in the optomechanical implementation of the $\Zthree$ Rabi model, we obtain a different phase for the qutrit term, $g HAH^{\dagger} = g(e^{-\pi i/6} Z + e^{\pi i/6} Z^{\dagger})$ [see Eqs.~(\ref{eq:optomechanical-rabi}) and (\ref{eq:optomechanical-magnetic-term})].

\section{Charge qubit for the \texorpdfstring{$\Zthree$}{Z3} Rabi model}
\label{app:charge-qubit}

\begin{figure}[t]
    \centering
    \subfloat[]{
         \centering
         \includegraphics[width=0.4\linewidth]{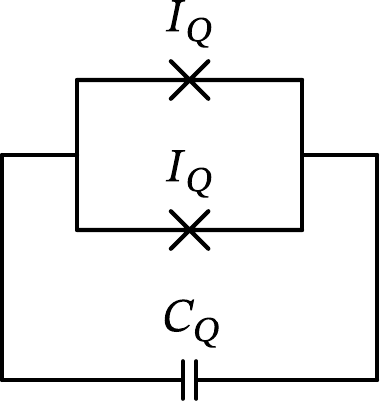}
         \vspace{0.1pt} 
         \label{fig:2nd-harmonic-qubit-a}}
     \hfill
    \subfloat[]{
         \centering
         \includegraphics[width=0.5\linewidth]{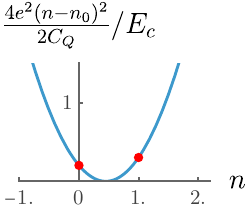}
         \label{fig:second-harmonic-qubit-b}}
    \caption{Second-harmonic CPB qubit  used for a superconducting-circuit realization of the $\Zthree$ Rabi model, as shown in Fig.~\ref{fig:superconducting-Rabi}. (a) The electric circuit of the second-harmonic CPB qubit. (b) The capacitor potential in Eq.~\eqref{eq:second-harmonic-qubit} for $n_0 = 0.45 = 0.5 - 0.05 $. The red dots indicate the eigenenergies of the second-harmonic CPB qubit.}
    \label{fig:second-harmonic-qubit}
\end{figure}

Here we provide details on the superconducting qubit architecture used in Sec.~\ref{sec:superconducting-implementation}. While the Cooper Pair Box (CPB) qubit is probably the most well-known type of a charge qubit, it is not suitable for our purposes because its eigenstates are not charge eigenstates but rather their symmetric/antisymmetric superpositions. In other words, in the charge basis the CPB Hamiltonian is proportional to $\sigma^x$. 
This is undesirable because in this case the Hamiltonian of the qubit chain does not commute with the total qubit-excitation-number operator $S^z$.

\begin{figure*}[th]
    \captionsetup[subfloat]{captionskip=-135pt} 
    \centering
    \subfloat[]{
         \centering
        \includegraphics[width=0.32\linewidth]{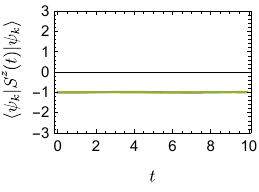}
         \label{fig:1st-spin-correlator}}
    \subfloat[]{
         \centering
        \includegraphics[width=0.274\linewidth]{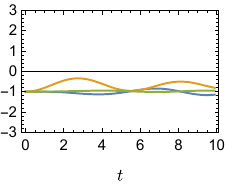}
         \label{fig:2nd-spin-correlator}}
    \subfloat[]{
         \centering
        \includegraphics[width=0.37\linewidth]{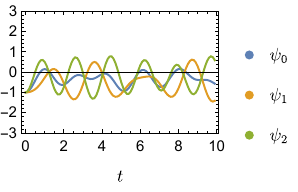}
         \label{fig:3rd-spin-correlator}}
    \caption{Time dependence of the $z$-component of the total spin operator in the presence of a symmetry-breaking disorder~(\ref{eq:qb-dis-hamiltonian}). In order to illustrate that the Rabi model implementation is robust, we plot $\langle S^z(t) \rangle$ for three disorder strengths. 
    We use normally distributed disorder $\Delta_j$ with zero mean and fixed standard deviation $\sigma$.  
    In panel a) $\sigma / \Omega_{\text{QB}} = 0.1$; panel b) $\sigma / \Omega_{\text{QB}} = 0.3$; panel c) $\sigma/ \Omega_{\text{QB}} = 1$.}
    \label{fig:spin-excitation-plot}
\end{figure*}

Therefore, we need some other type of charge qubit. Recall that in the CPB Hamiltonian, the $\sigma^x$ term arises from the Josephson junction (JJ). Thus, removing JJ one obtains a Hamiltonian proportional to $\sigma^z$ in the charge basis.
While this approach solves the issue with the $\Zthree$ symmetry, such a system cannot be called a qubit because of the absence of non-linearity.

Thus, we want two things simultaneously: (i) the Hamiltonian must be proportional to $\sigma^z$ in the charge basis, and (ii) in the qubit circuit there must be JJ that provides the non-linearity. The solution satisfying both requirements turns out to be a higher-harmonic Josephson junction. Usually it is neglected, but strictly speaking the JJ Hamiltonian always includes higher harmonics:
\begin{equation}
    H_{JJ} = E_{J,1} \cos{\hat \phi} + E_{J,2} \cos({2\hat \phi}) + \dots,
\end{equation}
where $\hat \phi$ is the superconducting phase and $E_{J,i} = I_{Q,i} \Phi_0 / (2\pi)$, with $\Phi_0 = 2\pi \hbar / 2e$ being the superconducting magnetic flux quantum and $I_{Q,i}$ being the critical current of the $i$th harmonic of the Josephson junction.

In practice, the higher harmonics are usually considerably weaker $E_{J,2} \ll E_{J,1}$. However, if we completely eliminate the first harmonic, the second harmonic becomes dominant. The second harmonic, on the one hand, can help to suppress the noise and, at the same time, does not induce any coupling in the qubit subspace:
\begin{equation}
\begin{aligned}
        \cos{2\hat \phi}\bigg|_{\mathcal{V}} &= \frac{1}{2}\left(e^{2i\hat\phi} + e^{-2i\hat \phi}\right)\bigg|_{\mathcal{V}} \\
        &= \frac{1}{2}\left((\sigma^+)^2 + (\sigma^-)^2\right) =0.
\end{aligned}
\end{equation}

In order to eliminate the first harmonic, we can use a Superconducting Quantum Interference Device (SQUID) \cite{valentini_parityconserving_2024}. If the magnetic flux through the SQUID is tuned to $\Phi = \pi$, then the SQUID Hamiltonian reads
\begin{equation}
\begin{aligned}
    &H_{\text{SQUID}} = E_{J,1} \cos(\hat \phi) + E_{J,2} \cos(2\hat \phi) \\ &+ E_{J,1} \cos(\hat \phi + \pi) 
    + E_{J,2} \cos(2\hat \phi + 2\pi) \\
    &= 2 E_{J,2} \cos(2\hat \phi),
\end{aligned}
\end{equation}
which has the desired form with only the second harmonic being present.

Consequently, using the second-harmonic JJ we can construct a charge qubit, which we refer to as the second-harmonic CPB qubit, described by the Hamiltonian (see Fig. \ref{fig:second-harmonic-qubit})
\begin{equation}
\label{eq:second-harmonic-qubit}
    H = \frac{4 e^2}{2C_{\text{Q}}} (\hat n - n_0)^2 + 2 E_{\text{J},2} \cos(2\hat\phi).
\end{equation}
Here, we have the capacitor energy $E_C = 2 e^2/C_Q$. By biasing the offset charge slightly away from the symmetric point ($n_0 = 0.5 - n_{\mathrm{off}}$ with $n_{\mathrm{off}} \ll 1$), the low-energy Hamiltonian of the qubit becomes
\begin{equation}
    H_{\text{Q}} = \frac{4 e^2 n_{\mathrm{off}}}{C_{\text{Q}}} \sigma^z.
\end{equation}
Hence, we obtain the qubit Hamiltonian proportional to $\sigma_z$, as desired. Here we assume that $\frac{4e^2}{2C_{\text{Q}}} \gg 2 E_{\text{J},2}$, i.e., the charging energy dominates over the Josephson energy, as is usually the case for a charge qubit.

To conclude, the suggested charge qubit consists of a capacitor and a second-harmonic JJ. One can think of it as a modified version of a Cooper Pair Box qubit. Here, we only briefly describe its blueprint to further support the implementation proposed in Sec.~\ref{sec:superconducting-implementation}.

\section{Effects of disorder in the superconducting implementation of the \texorpdfstring{$\Zthree$}{Z3} Rabi model}
\label{app:disorder}
\subsection{\texorpdfstring{$\mathrm{U}(1)$}{U(1)} symmetry-breaking disorder}

In this appendix, we discuss the effects of disorder on the qubit-boson ring~\eqref{eq:physical-hamiltonian}.
First, consider a $\mathrm{U}(1)$ symmetry-breaking term. It breaks the conservation of $\hat S^z$, the total number of qubit excitations. The $\mathrm{U}(1)$ symmetry is broken by a small misalignment of the Zeeman terms with the $\hat z$-direction: 
\begin{equation} \label{eq:qb-dis-hamiltonian}
    H_{\text{QB,dis}} = H_{\text{QB}} + \sum_{j=0}^2 \Delta_j \sigma^x_{j}.
\end{equation}
As explained in App. \ref{app:charge-qubit}, for the superconducting-circuit realization of the qubit-boson ring, the condition of vanishing disorder $\Delta_j = 0$ relies on the fine-tuning of the SQUID magnetic field. 
Therefore, in a realistic setting, a small detuning is inevitable $\Delta_j = I_{Q,1}\Phi_0/(2\pi)\sin(\Delta \Phi_j)$.

As a result, the qubit-boson ring Hamiltonian no longer commutes with the $S^z$ operator~\footnote{To not overcomplicate the notation, here we denote $S^z(t) = e^{iH_{\text{QB,dis}}t}S^z(k=0) e^{-iH_{\text{QB,dis}}t}$.}, $[H_{QB,\text{dis}}, S^z] \neq 0 $. This means that the exact mapping to the $\Zthree$ Rabi model derived in the main text does not hold anymore. However, assuming that the disorder is considerably weaker than other parameters in the Hamiltonian, we can still hope that the dynamics remains approximately that of the Rabi model. An analytic estimate is somewhat cumbersome, and we rely on numerical simulations of the qubit–boson ring. The results are shown in Fig.~\ref{fig:spin-excitation-plot}. The figure shows the time evolution of $\langle S^z(t)\rangle$ for various strengths of the disorder. The simulation indicates that for small disorder the evolution of the qubit-boson ring stays sufficiently close to the single excitation subspace $\mathcal{H}_1$, as $\langle S^z (t) \rangle\approx -1$ at all simulated times. We speculate that even when $S^z$ is not strictly conserved, the spins of the system collectively precess around the $S^z=-1$ direction, effectively keeping $\langle S^z \rangle$ near $-1$.

\subsection{\texorpdfstring{$\Zthree$}{Z3} symmetry-breaking disorder}

Now, assuming that conservation of $S^z$ is maintained ($\Delta_j=0$), we examine a different type of disorder: spatial variations in the system parameters. In an inhomogeneous qubit–boson ring, the $\Zthree$ symmetry is obviously broken. However, we want to investigate how this disorder manifests itself after the projection onto the single excitation subspace.
The disordered Hamiltonian in the full Hilbert space reads 
\begin{equation}
\label{eq:rabi-breaking-disorder}
\begin{aligned}
  H_{\text{QB}}' = &\sum_{j=0}^2 \epsilon_{j} \sigma_j^z + \sum_{j=0}^2 \hbar \Omega_{\text{QB},j} \hat a^{\dagger}_j \hat a_j  \\
  &+ \sum_{j=0}^2 g_{j,j+1} \left[\sigma_j^+ \sigma_{j+1}^- e^{i (\hat x_j - \hat x_{j+1})} +\text{H.c}\right],
\end{aligned}
\end{equation}
where we introduced spatial non-uniformity of the system parameters. The positional index $j$ is valued modulo $3$, $j+3 = j$.

In the superconducting implementation the couplings become $\epsilon_j = 4e^2n_{\mathrm{off},j} / (2 C_{Q,j})$, $\Omega_{\text{QB},j} = 1 / \sqrt{L_{\text{B},j}C_{\text{B},j}}$, and $g_{j,j+1} = I_{\text{R},j}\Phi_0/(4\pi)$,
where the subscript $j$ labels the parameters of the elements belonging to the $j$th branch of the circuit (see Fig.~\ref{fig:superconducting-Rabi}).

For convenience, we decompose the coupling parameters into a homogeneous part and a disorder: $\epsilon_j = \epsilon + \Delta \epsilon_j$, $\Omega_{\text{QB},j} = \Omega_{\text{QB}} + \Delta \Omega_{\text{QB},j}$, and $g_{j,j+1} = g + \Delta g_{j,j+1}$, where we assume that the spatial average of the disorder vanishes: $\sum_{j=0}^2 \Delta \epsilon_j =  \sum_{j=0}^2 \Delta \Omega_{\text{QB},j} = \sum_{j=0}^2 \Delta g_{j,j+1} =0$. This can always be achieved by adjusting the homogeneous parts of the couplings.

Then, after projecting to the single-excitation sector, the Hamiltonian~$H_{\text{QB}^{\prime}}$ reduces to
\begin{widetext}
\begin{align}
    H_{\text{R}}' &= H_{\text{R}} + \epsilon(2) X 
    +\epsilon(1)  X^{\dagger} + (X + X^{\dagger}) \sum\limits_{k=1}^2 \Delta g(k) Z^k
    + \hbar \Omega_{\text{QB}}(1) \sum\limits_{k=0}^2 \hat a^{\dagger}(k+1) \hat a(k) + \hbar \Omega_{\text{QB}}(2) \sum\limits_{k=0}^2 \hat a^{\dagger}(k) \hat a(k+1) \notag\\
    &+ \sum\limits_{l=1}^2 \frac{\eta \hbar \Omega_{\text{QB}}(l)}{2^{3/2}} \sum\limits_{k=0}^2 [\hat a(k+l) + \hat a^{\dagger}(-k-l)] Z^k,
\end{align}
\end{widetext} 
where the bosonic frequencies satify the following property, $\Omega_{\text{QB}}(1) = (\Omega_{\text{QB}}(2))^*$.
One can clearly see that, in the presence of disorder, the first bosonic mode is no longer decoupled. We expect that the $\Zthree$ cat state structure of the eigenstates should remain intact for sufficiently small disorder. However, a detailed investigation lies beyond the scope of this paper.

\bibliography{references}

@misc{aharonov_polynomial_2007,
  title = {Polynomial {{Quantum Algorithms}} for {{Additive}} Approximations of the {{Potts}} Model and Other {{Points}} of the {{Tutte Plane}}},
  author = {Aharonov, Dorit and Arad, Itai and Eban, Elad and Landau, Zeph},
  year = {2007},
  month = feb,
  number = {arXiv:quant-ph/0702008},
  eprint = {quant-ph/0702008},
  publisher = {arXiv},
  urldate = {2024-10-09},
  archiveprefix = {arXiv}
}

@article{albert_quantum_2012,
  title = {Quantum {{Rabi Model}} for {{N}} -{{State Atoms}}},
  author = {Albert, Victor V.},
  year = {2012},
  month = may,
  journal = {Phys. Rev. Lett.},
  volume = {108},
  number = {18},
  pages = {180401},
  issn = {0031-9007, 1079-7114},
  doi = {10.1103/PhysRevLett.108.180401},
  urldate = {2024-04-15},
  copyright = {http://link.aps.org/licenses/aps-default-license},
  langid = {english}
}

@article{allman_tunable_2014,
  title = {Tunable {{Resonant}} and {{Nonresonant Interactions}} between a {{Phase Qubit}} and \${{LC}}\$ {{Resonator}}},
  author = {Allman, M.S. and Whittaker, J.D. and {Castellanos-Beltran}, M. and Cicak, K. and {da Silva}, F. and DeFeo, M.P. and Lecocq, F. and Sirois, A. and Teufel, J.D. and Aumentado, J. and Simmonds, R.W.},
  year = {2014},
  month = mar,
  journal = {Phys. Rev. Lett.},
  volume = {112},
  number = {12},
  pages = {123601},
  publisher = {American Physical Society},
  doi = {10.1103/PhysRevLett.112.123601},
  urldate = {2024-10-10}
}

@article{astafiev_resonance_2010,
  title = {Resonance {{Fluorescence}} of a {{Single Artificial Atom}}},
  author = {Astafiev, O. and Zagoskin, A. M. and Abdumalikov, A. A. and Pashkin, {\relax Yu}. A. and Yamamoto, T. and Inomata, K. and Nakamura, Y. and Tsai, J. S.},
  year = {2010},
  month = feb,
  journal = {Science},
  volume = {327},
  number = {5967},
  pages = {840--843},
  publisher = {American Association for the Advancement of Science},
  doi = {10.1126/science.1181918},
  urldate = {2025-08-18}
}

@article{baxter_critical_1982,
  title = {Critical {{Antiferromagnetic Square-Lattice Potts Model}}},
  author = {Baxter, R. J.},
  year = {1982},
  journal = {Proc. R. Soc. Lond. Ser. Math. Phys. Sci.},
  volume = {383},
  number = {1784},
  eprinttype = {jstor},
  pages = {43--54},
  publisher = {Royal Society},
  issn = {0080-4630},
  urldate = {2024-10-09}
}

@article{bazavan_synthetic_2024,
  title = {Synthetic \${\textbackslash}mathbb\{\vphantom\}{{Z}}\vphantom\{\}\_2\$ Gauge Theories Based on Parametric Excitations of Trapped Ions},
  author = {B{\u a}z{\u a}van, O. and Saner, S. and Tirrito, E. and Araneda, G. and Srinivas, R. and Bermudez, A.},
  year = {2024},
  month = jul,
  journal = {Commun Phys},
  volume = {7},
  number = {1},
  primaryclass = {quant-ph},
  pages = {229},
  issn = {2399-3650},
  urldate = {2025-08-24},
  archiveprefix = {arXiv}
}

@article{bermudez_synthetic_2011,
  title = {Synthetic {{Gauge Fields}} for {{Vibrational Excitations}} of {{Trapped}} Ions},
  author = {Bermudez, A. and Schaetz, T. and Porras, D.},
  year = {2011},
  month = oct,
  journal = {Phys. Rev. Lett.},
  volume = {107},
  number = {15},
  primaryclass = {quant-ph},
  pages = {150501},
  issn = {0031-9007, 1079-7114},
  urldate = {2024-04-16},
  archiveprefix = {arXiv}
}

@article{bernien_probing_2017,
  title = {Probing Many-Body Dynamics on a 51-Atom Quantum Simulator},
  author = {Bernien, Hannes and Schwartz, Sylvain and Keesling, Alexander and Levine, Harry and Omran, Ahmed and Pichler, Hannes and Choi, Soonwon and Zibrov, Alexander S. and Endres, Manuel and Greiner, Markus and Vuleti{\'c}, Vladan and Lukin, Mikhail D.},
  year = {2017},
  month = nov,
  journal = {Nature},
  volume = {551},
  number = {7682},
  pages = {579--584},
  publisher = {Nature Publishing Group},
  issn = {1476-4687},
  doi = {10.1038/nature24622},
  urldate = {2025-08-12},
  copyright = {2017 Macmillan Publishers Limited, part of Springer Nature. All rights reserved.},
  langid = {english}
}

@article{bosco_fully_2022,
  title = {Fully {{Tunable Longitudinal Spin-Photon Interactions}} in {{Si}} and {{Ge Quantum Dots}}},
  author = {Bosco, Stefano and Scarlino, Pasquale and Klinovaja, Jelena and Loss, Daniel},
  year = {2022},
  month = aug,
  journal = {Phys. Rev. Lett.},
  volume = {129},
  number = {6},
  pages = {066801},
  issn = {0031-9007, 1079-7114},
  doi = {10.1103/PhysRevLett.129.066801},
  urldate = {2024-08-28},
  langid = {english}
}

@article{bouchiat_quantum_1998,
  title = {Quantum Coherence with a Single {{Cooper}} Pair},
  author = {Bouchiat, V. and Vion, D. and Joyez, P. and Esteve, D. and Devoret, M. H.},
  year = {1998},
  month = jan,
  journal = {Phys. Scr.},
  volume = {1998},
  number = {T76},
  pages = {165},
  publisher = {IOP Publishing},
  issn = {1402-4896},
  doi = {10.1238/Physica.Topical.076a00165},
  urldate = {2024-10-10},
  langid = {english}
}

@article{braak_integrability_2011,
  title = {Integrability of the {{Rabi Model}}},
  author = {Braak, D.},
  year = {2011},
  month = aug,
  journal = {Phys. Rev. Lett.},
  volume = {107},
  number = {10},
  pages = {100401},
  publisher = {American Physical Society},
  doi = {10.1103/PhysRevLett.107.100401},
  urldate = {2024-02-06}
}

@article{braumuller_analog_2017,
  title = {Analog Quantum Simulation of the {{Rabi}} Model in the Ultra-Strong Coupling Regime},
  author = {Braum{\"u}ller, Jochen and Marthaler, Michael and Schneider, Andre and Stehli, Alexander and Rotzinger, Hannes and Weides, Martin and Ustinov, Alexey V.},
  year = {2017},
  month = oct,
  journal = {Nat Commun},
  volume = {8},
  number = {1},
  pages = {779},
  publisher = {Nature Publishing Group},
  issn = {2041-1723},
  doi = {10.1038/s41467-017-00894-w},
  urldate = {2024-09-04},
  copyright = {2017 The Author(s)},
  langid = {english}
}

@article{bravyi_schrieffer_2011,
  title = {Schrieffer--{{Wolff}} Transformation for Quantum Many-Body Systems},
  author = {Bravyi, Sergey and DiVincenzo, David P. and Loss, Daniel},
  year = {2011},
  month = oct,
  journal = {Annals of Physics},
  volume = {326},
  number = {10},
  pages = {2793--2826},
  issn = {0003-4916},
  doi = {10.1016/j.aop.2011.06.004},
  urldate = {2025-05-02}
}

@article{brydges_probing_2019,
  title = {Probing {{R{\'e}nyi}} Entanglement Entropy via Randomized Measurements},
  author = {Brydges, Tiff and Elben, Andreas and Jurcevic, Petar and Vermersch, Beno{\^i}t and Maier, Christine and Lanyon, Ben P. and Zoller, Peter and Blatt, Rainer and Roos, Christian F.},
  year = {2019},
  month = apr,
  journal = {Science},
  volume = {364},
  number = {6437},
  pages = {260--263},
  publisher = {American Association for the Advancement of Science},
  doi = {10.1126/science.aau4963},
  urldate = {2025-08-12}
}

@article{cao_parametrically_2024,
  title = {Parametrically Controlled Chiral Interface for Superconducting Quantum Devices},
  author = {Cao, Xi and Irfan, Abdullah and Mollenhauer, Michael and Singirikonda, Kaushik and Pfaff, Wolfgang},
  year = {2024},
  month = dec,
  journal = {Phys. Rev. Appl.},
  volume = {22},
  number = {6},
  pages = {064023},
  publisher = {American Physical Society},
  doi = {10.1103/PhysRevApplied.22.064023},
  urldate = {2025-08-24}
}

@article{chen_shortcuts_2021,
  title = {Shortcuts to {{Adiabaticity}} for the {{Quantum Rabi Model}}: {{Efficient Generation}} of {{Giant Entangled Cat States}} via {{Parametric Amplification}}},
  shorttitle = {Shortcuts to {{Adiabaticity}} for the {{Quantum Rabi Model}}},
  author = {Chen, Ye-Hong and Qin, Wei and Wang, Xin and Miranowicz, Adam and Nori, Franco},
  year = {2021},
  month = jan,
  journal = {Phys. Rev. Lett.},
  volume = {126},
  number = {2},
  pages = {023602},
  publisher = {American Physical Society},
  doi = {10.1103/PhysRevLett.126.023602},
  urldate = {2024-08-20}
}

@article{chen_singlephotondriven_2017,
  title = {Single-Photon-Driven High-Order Sideband Transitions in an Ultrastrongly Coupled Circuit-Quantum-Electrodynamics System},
  author = {Chen, Zhen and Wang, Yimin and Li, Tiefu and Tian, Lin and Qiu, Yueyin and Inomata, Kunihiro and Yoshihara, Fumiki and Han, Siyuan and Nori, Franco and Tsai, J. S. and You, J. Q.},
  year = {2017},
  month = jul,
  journal = {Phys. Rev. A},
  volume = {96},
  number = {1},
  pages = {012325},
  issn = {2469-9926, 2469-9934},
  doi = {10.1103/PhysRevA.96.012325},
  urldate = {2024-08-28},
  copyright = {http://link.aps.org/licenses/aps-default-license},
  langid = {english}
}

@article{dehollain_nagaoka_2020,
  title = {Nagaoka Ferromagnetism Observed in a Quantum Dot Plaquette},
  author = {Dehollain, J. P. and Mukhopadhyay, U. and Michal, V. P. and Wang, Y. and Wunsch, B. and Reichl, C. and Wegscheider, W. and Rudner, M. S. and Demler, E. and Vandersypen, L. M. K.},
  year = {2020},
  month = mar,
  journal = {Nature},
  volume = {579},
  number = {7800},
  pages = {528--533},
  publisher = {Nature Publishing Group},
  issn = {1476-4687},
  doi = {10.1038/s41586-020-2051-0},
  urldate = {2025-08-12},
  copyright = {2020 The Author(s), under exclusive licence to Springer Nature Limited},
  langid = {english}
}

@article{felicetti_quantum_2017,
  title = {Quantum {{Rabi}} Model in a Superfluid {{Bose-Einstein}} Condensate},
  author = {Felicetti, S. and Romero, G. and Solano, E. and Sab{\'i}n, C.},
  year = {2017},
  month = sep,
  journal = {Phys. Rev. A},
  volume = {96},
  number = {3},
  pages = {033839},
  issn = {2469-9926, 2469-9934},
  doi = {10.1103/PhysRevA.96.033839},
  urldate = {2024-08-28},
  copyright = {https://link.aps.org/licenses/aps-default-license},
  langid = {english}
}

@article{fendley_parafermionic_2012,
  title = {Parafermionic Edge Zero Modes in {{Zn-invariant}} Spin Chains},
  author = {Fendley, Paul},
  year = {2012},
  month = nov,
  journal = {J. Stat. Mech.},
  volume = {2012},
  number = {11},
  pages = {P11020},
  publisher = {{IOP Publishing and SISSA}},
  issn = {1742-5468},
  doi = {10.1088/1742-5468/2012/11/P11020},
  urldate = {2023-04-04},
  langid = {english}
}

@article{forn-diaz_ultrastrong_2017,
  title = {Ultrastrong Coupling of a Single Artificial Atom to an Electromagnetic Continuum in the Nonperturbative Regime},
  author = {{Forn-D{\'i}az}, P. and {Garc{\'i}a-Ripoll}, J. J. and Peropadre, B. and Orgiazzi, J.-L. and Yurtalan, M. A. and Belyansky, R. and Wilson, C. M. and Lupascu, A.},
  year = {2017},
  month = jan,
  journal = {Nature Phys},
  volume = {13},
  number = {1},
  pages = {39--43},
  publisher = {Nature Publishing Group},
  issn = {1745-2481},
  doi = {10.1038/nphys3905},
  urldate = {2024-08-28},
  copyright = {2016 Springer Nature Limited},
  langid = {english}
}

@article{fradkin_disorder_1980,
  title = {Disorder Variables and Para-Fermions in Two-Dimensional Statistical Mechanics},
  author = {Fradkin, Eduardo and Kadanoff, Leo P.},
  year = {1980},
  month = aug,
  journal = {Nuclear Physics B},
  volume = {170},
  number = {1},
  pages = {1--15},
  issn = {0550-3213},
  doi = {10.1016/0550-3213(80)90472-1},
  urldate = {2024-08-20}
}

@article{gu_microwave_2017,
  title = {Microwave Photonics with Superconducting Quantum Circuits},
  author = {Gu, Xiu and Kockum, Anton Frisk and Miranowicz, Adam and Liu, Yu-xi and Nori, Franco},
  year = {2017},
  month = nov,
  journal = {Physics Reports},
  series = {Microwave Photonics with Superconducting Quantum Circuits},
  volume = {718--719},
  pages = {1--102},
  issn = {0370-1573},
  doi = {10.1016/j.physrep.2017.10.002},
  urldate = {2025-08-24}
}

@article{hensgens_quantum_2017,
  title = {Quantum Simulation of a {{Fermi}}--{{Hubbard}} Model Using a Semiconductor Quantum Dot Array},
  author = {Hensgens, T. and Fujita, T. and Janssen, L. and Li, Xiao and Van Diepen, C. J. and Reichl, C. and Wegscheider, W. and Das Sarma, S. and Vandersypen, L. M. K.},
  year = {2017},
  month = aug,
  journal = {Nature},
  volume = {548},
  number = {7665},
  pages = {70--73},
  publisher = {Nature Publishing Group},
  issn = {1476-4687},
  doi = {10.1038/nature23022},
  urldate = {2025-08-12},
  copyright = {2017 Macmillan Publishers Limited, part of Springer Nature. All rights reserved.},
  langid = {english}
}

@article{hu_controllable_2007,
  title = {Controllable Coupling of Superconducting Transmission-Line Resonators},
  author = {Hu, Yong and Xiao, Yun-Feng and Zhou, Zheng-Wei and Guo, Guang-Can},
  year = {2007},
  month = jan,
  journal = {Phys. Rev. A},
  volume = {75},
  number = {1},
  pages = {012314},
  publisher = {American Physical Society},
  doi = {10.1103/PhysRevA.75.012314},
  urldate = {2024-10-10}
}

@article{hwang_largescale_2013,
  title = {Large-Scale Maximal Entanglement and {{Majorana}} Bound States in Coupled Circuit Quantum Electrodynamic Systems},
  author = {Hwang, Myung-Joong and Choi, Mahn-Soo},
  year = {2013},
  month = mar,
  journal = {Phys. Rev. B},
  volume = {87},
  number = {12},
  pages = {125404},
  issn = {1098-0121, 1550-235X},
  doi = {10.1103/PhysRevB.87.125404},
  urldate = {2024-04-04},
  copyright = {http://link.aps.org/licenses/aps-default-license},
  langid = {english}
}

@article{hwang_quantum_2015,
  title = {Quantum {{Phase Transition}} and {{Universal Dynamics}} in the {{Rabi Model}}},
  author = {Hwang, Myung-Joong and Puebla, Ricardo and Plenio, Martin B.},
  year = {2015},
  month = oct,
  journal = {Phys. Rev. Lett.},
  volume = {115},
  number = {18},
  pages = {180404},
  publisher = {American Physical Society},
  doi = {10.1103/PhysRevLett.115.180404},
  urldate = {2024-02-06}
}

@article{kapit_quantum_2013,
  title = {Quantum Simulation Architecture for Lattice Bosons in Arbitrary, Tunable, External Gauge Fields},
  author = {Kapit, Eliot},
  year = {2013},
  month = jun,
  journal = {Phys. Rev. A},
  volume = {87},
  number = {6},
  pages = {062336},
  publisher = {American Physical Society},
  doi = {10.1103/PhysRevA.87.062336},
  urldate = {2025-08-24}
}

@article{kiczynski_engineering_2022,
  title = {Engineering Topological States in Atom-Based Semiconductor Quantum Dots},
  author = {Kiczynski, M. and Gorman, S. K. and Geng, H. and Donnelly, M. B. and Chung, Y. and He, Y. and Keizer, J. G. and Simmons, M. Y.},
  year = {2022},
  month = jun,
  journal = {Nature},
  volume = {606},
  number = {7915},
  pages = {694--699},
  publisher = {Nature Publishing Group},
  issn = {1476-4687},
  doi = {10.1038/s41586-022-04706-0},
  urldate = {2025-08-12},
  copyright = {2022 The Author(s)},
  langid = {english}
}

@article{kiefer_floquetengineered_2019,
  title = {Floquet-Engineered Vibrational Dynamics in a Two-Dimensional Array of Trapped Ions},
  author = {Kiefer, Philip and Hakelberg, Frederick and Wittemer, Matthias and Berm{\'u}dez, Alejandro and Porras, Diego and Warring, Ulrich and Schaetz, Tobias},
  year = {2019},
  month = nov,
  journal = {Phys. Rev. Lett.},
  volume = {123},
  number = {21},
  primaryclass = {quant-ph},
  pages = {213605},
  issn = {0031-9007, 1079-7114},
  urldate = {2025-08-24},
  archiveprefix = {arXiv}
}

@article{kitaev_unpaired_2001,
  title = {Unpaired {{Majorana}} Fermions in Quantum Wires},
  author = {Kitaev, A. Yu},
  year = {2001},
  month = oct,
  journal = {Phys.-Usp.},
  volume = {44},
  number = {10S},
  pages = {131},
  issn = {1063-7869},
  doi = {10.1070/1063-7869/44/10S/S29},
  urldate = {2025-08-24},
  langid = {english}
}

@article{kozin_cavityenhanced_2025,
  title = {Cavity-Enhanced Superconductivity via Band Engineering},
  author = {Kozin, Valerii K. and Thingstad, Even and Loss, Daniel and Klinovaja, Jelena},
  year = {2025},
  month = jan,
  journal = {Phys. Rev. B},
  volume = {111},
  number = {3},
  pages = {035410},
  publisher = {American Physical Society},
  doi = {10.1103/PhysRevB.111.035410},
  urldate = {2025-06-05}
}

@article{kozin_quantum_2024,
  title = {Quantum Phase Transitions and Cat States in Cavity-Coupled Quantum Dots},
  author = {Kozin, Valerii K. and Miserev, Dmitry and Loss, Daniel and Klinovaja, Jelena},
  year = {2024},
  month = aug,
  journal = {Phys. Rev. Res.},
  volume = {6},
  number = {3},
  pages = {033188},
  publisher = {American Physical Society},
  doi = {10.1103/PhysRevResearch.6.033188},
  urldate = {2025-05-02}
}

@article{kurpiers_deterministic_2018,
  title = {Deterministic Quantum State Transfer and Remote Entanglement Using Microwave Photons},
  author = {Kurpiers, P. and Magnard, P. and Walter, T. and Royer, B. and Pechal, M. and Heinsoo, J. and Salath{\'e}, Y. and Akin, A. and Storz, S. and Besse, J.-C. and Gasparinetti, S. and Blais, A. and Wallraff, A.},
  year = {2018},
  month = jun,
  journal = {Nature},
  volume = {558},
  number = {7709},
  pages = {264--267},
  publisher = {Nature Publishing Group},
  issn = {1476-4687},
  doi = {10.1038/s41586-018-0195-y},
  urldate = {2025-08-18},
  copyright = {2018 Macmillan Publishers Ltd., part of Springer Nature},
  langid = {english}
}

@article{kyprianidis_observation_2021,
  title = {Observation of a Prethermal Discrete Time Crystal},
  author = {Kyprianidis, A. and Machado, F. and Morong, W. and Becker, P. and Collins, K. S. and Else, D. V. and Feng, L. and Hess, P. W. and Nayak, C. and Pagano, G. and Yao, N. Y. and Monroe, C.},
  year = {2021},
  month = jun,
  journal = {Science},
  volume = {372},
  number = {6547},
  pages = {1192--1196},
  publisher = {American Association for the Advancement of Science},
  doi = {10.1126/science.abg8102},
  urldate = {2025-08-12}
}

@article{lehnert_measurement_2003,
  title = {Measurement of the {{Excited-State Lifetime}} of a {{Microelectronic Circuit}}},
  author = {Lehnert, K. W. and Bladh, K. and Spietz, L. F. and Gunnarsson, D. and Schuster, D. I. and Delsing, P. and Schoelkopf, R. J.},
  year = {2003},
  month = jan,
  journal = {Phys. Rev. Lett.},
  volume = {90},
  number = {2},
  pages = {027002},
  publisher = {American Physical Society},
  doi = {10.1103/PhysRevLett.90.027002},
  urldate = {2024-10-10}
}

@misc{lotkov_cat_2025b,
  title = {Cat States in One- and Two-Mode \${\textbackslash}mathbb\{\vphantom\}{{Z}}\vphantom\{\}\_3\$ {{Rabi}} Models},
  author = {Lotkov, Anatoliy I. and Kurlov, Denis V. and Kozin, Valerii K. and Klinovaja, Jelena and Loss, Daniel},
  year = {2025},
  month = sep,
  number = {arXiv:2509.08603},
  eprint = {2509.08603},
  primaryclass = {quant-ph},
  publisher = {arXiv},
  urldate = {2025-09-29},
  archiveprefix = {arXiv}
}

@article{madjarov_highfidelity_2020,
  title = {High-Fidelity Entanglement and Detection of Alkaline-Earth {{Rydberg}} Atoms},
  author = {Madjarov, Ivaylo S. and Covey, Jacob P. and Shaw, Adam L. and Choi, Joonhee and Kale, Anant and Cooper, Alexandre and Pichler, Hannes and Schkolnik, Vladimir and Williams, Jason R. and Endres, Manuel},
  year = {2020},
  month = aug,
  journal = {Nat. Phys.},
  volume = {16},
  number = {8},
  pages = {857--861},
  publisher = {Nature Publishing Group},
  issn = {1745-2481},
  doi = {10.1038/s41567-020-0903-z},
  urldate = {2025-08-12},
  copyright = {2020 The Author(s), under exclusive licence to Springer Nature Limited},
  langid = {english}
}

@article{makhlin_quantumstate_2001,
  title = {Quantum-State Engineering with {{Josephson-junction}} Devices},
  author = {Makhlin, Yuriy and Sch{\"o}n, Gerd and Shnirman, Alexander},
  year = {2001},
  month = may,
  journal = {Rev. Mod. Phys.},
  volume = {73},
  number = {2},
  pages = {357--400},
  publisher = {American Physical Society},
  doi = {10.1103/RevModPhys.73.357},
  urldate = {2024-10-10}
}

@article{manovitz_quantum_2025,
  title = {Quantum Coarsening and Collective Dynamics on a Programmable Simulator},
  author = {Manovitz, Tom and Li, Sophie H. and Ebadi, Sepehr and Samajdar, Rhine and Geim, Alexandra A. and Evered, Simon J. and Bluvstein, Dolev and Zhou, Hengyun and Koyluoglu, Nazli Ugur and Feldmeier, Johannes and Dolgirev, Pavel E. and Maskara, Nishad and Kalinowski, Marcin and Sachdev, Subir and Huse, David A. and Greiner, Markus and Vuleti{\'c}, Vladan and Lukin, Mikhail D.},
  year = {2025},
  month = feb,
  journal = {Nature},
  volume = {638},
  number = {8049},
  pages = {86--92},
  publisher = {Nature Publishing Group},
  issn = {1476-4687},
  doi = {10.1038/s41586-024-08353-5},
  urldate = {2025-08-12},
  copyright = {2025 The Author(s)},
  langid = {english}
}

@article{martinez_realtime_2016,
  title = {Real-Time Dynamics of Lattice Gauge Theories with a Few-Qubit Quantum Computer},
  author = {Martinez, Esteban A. and Muschik, Christine A. and Schindler, Philipp and Nigg, Daniel and Erhard, Alexander and Heyl, Markus and Hauke, Philipp and Dalmonte, Marcello and Monz, Thomas and Zoller, Peter and Blatt, Rainer},
  year = {2016},
  month = jun,
  journal = {Nature},
  volume = {534},
  number = {7608},
  pages = {516--519},
  publisher = {Nature Publishing Group},
  issn = {1476-4687},
  doi = {10.1038/nature18318},
  urldate = {2025-08-12},
  copyright = {2016 Springer Nature Limited},
  langid = {english}
}

@article{meth_simulating_2025,
  title = {Simulating Two-Dimensional Lattice Gauge Theories on a Qudit Quantum Computer},
  author = {Meth, Michael and Zhang, Jinglei and Haase, Jan F. and Edmunds, Claire and Postler, Lukas and Jena, Andrew J. and Steiner, Alex and Dellantonio, Luca and Blatt, Rainer and Zoller, Peter and Monz, Thomas and Schindler, Philipp and Muschik, Christine and Ringbauer, Martin},
  year = {2025},
  month = apr,
  journal = {Nat. Phys.},
  volume = {21},
  number = {4},
  pages = {570--576},
  publisher = {Nature Publishing Group},
  issn = {1745-2481},
  doi = {10.1038/s41567-025-02797-w},
  urldate = {2025-08-12},
  copyright = {2025 The Author(s)},
  langid = {english}
}

@article{nakamura_coherent_1999,
  title = {Coherent Control of Macroscopic Quantum States in a Single-{{Cooper-pair}} Box},
  author = {Nakamura, Y. and Pashkin, Yu A. and Tsai, J. S.},
  year = {1999},
  month = apr,
  journal = {Nature},
  volume = {398},
  number = {6730},
  pages = {786--788},
  publisher = {Nature Publishing Group},
  issn = {1476-4687},
  doi = {10.1038/19718},
  urldate = {2024-10-10},
  copyright = {1999 Macmillan Magazines Ltd.},
  langid = {english}
}

@article{niemczyk_circuit_2010,
  title = {Circuit Quantum Electrodynamics in the Ultrastrong-Coupling Regime},
  author = {Niemczyk, T. and Deppe, F. and Huebl, H. and Menzel, E. P. and Hocke, F. and Schwarz, M. J. and {Garcia-Ripoll}, J. J. and Zueco, D. and H{\"u}mmer, T. and Solano, E. and Marx, A. and Gross, R.},
  year = {2010},
  month = oct,
  journal = {Nature Phys},
  volume = {6},
  number = {10},
  pages = {772--776},
  publisher = {Nature Publishing Group},
  issn = {1745-2481},
  doi = {10.1038/nphys1730},
  urldate = {2024-08-28},
  copyright = {2010 Springer Nature Limited},
  langid = {english}
}

@misc{okada_efficient_2019,
  title = {Efficient Quantum and Simulated Annealing of {{Potts}} Models Using a Half-Hot Constraint},
  author = {Okada, Shuntaro and Ohzeki, Masayuki and Tanaka, Kazuyuki},
  year = {2019},
  month = apr,
  journal = {arXiv.org},
  doi = {10.7566/JPSJ.89.094801},
  urldate = {2024-10-09},
  langid = {english}
}

@article{rasmussen_controllable_2019,
  title = {Controllable Two-Qubit Swapping Gate Using Superconducting Circuits},
  author = {Rasmussen, S. E. and Christensen, K. S. and Zinner, N. T.},
  year = {2019},
  month = apr,
  journal = {Phys. Rev. B},
  volume = {99},
  number = {13},
  pages = {134508},
  publisher = {American Physical Society},
  doi = {10.1103/PhysRevB.99.134508},
  urldate = {2024-10-10}
}

@article{ricco_reshaping_2022,
  title = {Reshaping the {{Jaynes-Cummings}} Ladder with {{Majorana}} Bound States},
  author = {Ricco, L. S. and Kozin, V. K. and Seridonio, A. C. and Shelykh, I. A.},
  year = {2022},
  month = aug,
  journal = {Phys. Rev. A},
  volume = {106},
  number = {2},
  pages = {023702},
  publisher = {American Physical Society},
  doi = {10.1103/PhysRevA.106.023702},
  urldate = {2025-08-12}
}

@article{roushan_chiral_2017,
  title = {Chiral Ground-State Currents of Interacting Photons in a Synthetic Magnetic Field},
  author = {Roushan, P. and Neill, C. and Megrant, A. and Chen, Y. and Babbush, R. and Barends, R. and Campbell, B. and Chen, Z. and Chiaro, B. and Dunsworth, A. and Fowler, A. and Jeffrey, E. and Kelly, J. and Lucero, E. and Mutus, J. and O'Malley, P. J. J. and Neeley, M. and Quintana, C. and Sank, D. and Vainsencher, A. and Wenner, J. and White, T. and Kapit, E. and Neven, H. and Martinis, J.},
  year = {2017},
  month = feb,
  journal = {Nature Phys},
  volume = {13},
  number = {2},
  pages = {146--151},
  publisher = {Nature Publishing Group},
  issn = {1745-2481},
  doi = {10.1038/nphys3930},
  urldate = {2025-08-22},
  copyright = {2016 Springer Nature Limited},
  langid = {english}
}

@article{schneider_experimental_2012,
  title = {Experimental Quantum Simulations of Many-Body Physics with Trapped Ions},
  author = {Schneider, Ch and Porras, Diego and Schaetz, Tobias},
  year = {2012},
  month = feb,
  journal = {Rep. Prog. Phys.},
  volume = {75},
  number = {2},
  pages = {024401},
  issn = {0034-4885, 1361-6633},
  doi = {10.1088/0034-4885/75/2/024401},
  urldate = {2024-10-24},
  langid = {english}
}

@article{scholl_quantum_2021,
  title = {Quantum Simulation of {{2D}} Antiferromagnets with Hundreds of {{Rydberg}} Atoms},
  author = {Scholl, Pascal and Schuler, Michael and Williams, Hannah J. and Eberharter, Alexander A. and Barredo, Daniel and Schymik, Kai-Niklas and Lienhard, Vincent and Henry, Louis-Paul and Lang, Thomas C. and Lahaye, Thierry and L{\"a}uchli, Andreas M. and Browaeys, Antoine},
  year = {2021},
  month = jul,
  journal = {Nature},
  volume = {595},
  number = {7866},
  pages = {233--238},
  publisher = {Nature Publishing Group},
  issn = {1476-4687},
  doi = {10.1038/s41586-021-03585-1},
  urldate = {2025-08-12},
  copyright = {2021 The Author(s), under exclusive licence to Springer Nature Limited},
  langid = {english}
}

@article{sedov_chiral_2020,
  title = {Chiral {{Waveguide Optomechanics}}: {{First Order Quantum Phase Transitions}} with \$\{{\textbackslash}mathbb\{\vphantom{\}\}}{{Z}}\vphantom\{\}\vphantom\{\}\_\{3\}\$ {{Symmetry Breaking}}},
  shorttitle = {Chiral {{Waveguide Optomechanics}}},
  author = {Sedov, D.D. and Kozin, V.K. and Iorsh, I.V.},
  year = {2020},
  month = dec,
  journal = {Phys. Rev. Lett.},
  volume = {125},
  number = {26},
  pages = {263606},
  publisher = {American Physical Society},
  doi = {10.1103/PhysRevLett.125.263606},
  urldate = {2024-03-01}
}

@article{semeghini_probing_2021,
  title = {Probing Topological Spin Liquids on a Programmable Quantum Simulator},
  author = {Semeghini, G. and Levine, H. and Keesling, A. and Ebadi, S. and Wang, T. T. and Bluvstein, D. and Verresen, R. and Pichler, H. and Kalinowski, M. and Samajdar, R. and Omran, A. and Sachdev, S. and Vishwanath, A. and Greiner, M. and Vuleti{\'c}, V. and Lukin, M. D.},
  year = {2021},
  month = dec,
  journal = {Science},
  volume = {374},
  number = {6572},
  pages = {1242--1247},
  publisher = {American Association for the Advancement of Science},
  doi = {10.1126/science.abi8794},
  urldate = {2025-08-12}
}

@article{shafranjuk_twoqubit_2006,
  title = {Two-Qubit Gate Based on a Multiterminal Double-Barrier {{Josephson}} Junction},
  author = {Shafranjuk, S. E.},
  year = {2006},
  month = jul,
  journal = {Phys. Rev. B},
  volume = {74},
  number = {2},
  pages = {024521},
  publisher = {American Physical Society},
  doi = {10.1103/PhysRevB.74.024521},
  urldate = {2024-10-10}
}

@article{siewert_aspects_2000,
  title = {Aspects of {{Qubit Dynamics}} in the {{Presence}} of {{Leakage}}},
  author = {Siewert, Jens and Fazio, Rosario and Palma, G. Massimo and Sciacca, Emilio},
  year = {2000},
  month = mar,
  journal = {Journal of Low Temperature Physics},
  volume = {118},
  number = {5},
  pages = {795--804},
  issn = {1573-7357},
  doi = {10.1023/A:1004612016347},
  urldate = {2024-05-16},
  langid = {english}
}

@article{skogvoll_tunable_2021,
  title = {Tunable {{Anisotropic Quantum Rabi Model}} via a {{Magnon}}--{{Spin-Qubit Ensemble}}},
  author = {Skogvoll, Ida C. and Lidal, Jonas and Danon, Jeroen and Kamra, Akashdeep},
  year = {2021},
  month = dec,
  journal = {Phys. Rev. Applied},
  volume = {16},
  number = {6},
  pages = {064008},
  issn = {2331-7019},
  doi = {10.1103/PhysRevApplied.16.064008},
  urldate = {2024-08-28},
  langid = {english}
}

@article{sliwa_reconfigurable_2015,
  title = {Reconfigurable {{Josephson Circulator}}/{{Directional Amplifier}}},
  author = {Sliwa, K. M. and Hatridge, M. and Narla, A. and Shankar, S. and Frunzio, L. and Schoelkopf, R. J. and Devoret, M. H.},
  year = {2015},
  month = nov,
  journal = {Phys. Rev. X},
  volume = {5},
  number = {4},
  pages = {041020},
  publisher = {American Physical Society},
  doi = {10.1103/PhysRevX.5.041020},
  urldate = {2025-08-24}
}

@article{tan_domainwall_2021,
  title = {Domain-Wall Confinement and Dynamics in a Quantum Simulator},
  author = {Tan, W. L. and Becker, P. and Liu, F. and Pagano, G. and Collins, K. S. and De, A. and Feng, L. and Kaplan, H. B. and Kyprianidis, A. and Lundgren, R. and Morong, W. and Whitsitt, S. and Gorshkov, A. V. and Monroe, C.},
  year = {2021},
  month = jun,
  journal = {Nat. Phys.},
  volume = {17},
  number = {6},
  pages = {742--747},
  publisher = {Nature Publishing Group},
  issn = {1745-2481},
  doi = {10.1038/s41567-021-01194-3},
  urldate = {2025-08-12},
  copyright = {2021 The Author(s), under exclusive licence to Springer Nature Limited part of Springer Nature},
  langid = {english}
}

@phdthesis{timm_dynamics_2023,
  type = {{{DoctoralThesis}}},
  title = {Dynamics of Ion {{Coulomb}} Crystals},
  author = {Timm, Lars},
  year = {2023},
  doi = {10.15488/16052},
  urldate = {2024-10-24},
  copyright = {CC BY 3.0 DE},
  langid = {english},
  school = {Hannover : Institutionelles Repositorium der Leibniz Universit{\"a}t Hannover}
}

@article{valentini_parityconserving_2024,
  title = {Parity-Conserving {{Cooper-pair}} Transport and Ideal Superconducting Diode in Planar Germanium},
  author = {Valentini, Marco and Sagi, Oliver and Baghumyan, Levon and {de Gijsel}, Thijs and Jung, Jason and Calcaterra, Stefano and Ballabio, Andrea and Aguilera Servin, Juan and Aggarwal, Kushagra and Janik, Marian and Adletzberger, Thomas and Seoane Souto, Rub{\'e}n and Leijnse, Martin and Danon, Jeroen and Schrade, Constantin and Bakkers, Erik and Chrastina, Daniel and Isella, Giovanni and Katsaros, Georgios},
  year = {2024},
  month = jan,
  journal = {Nat Commun},
  volume = {15},
  number = {1},
  pages = {169},
  publisher = {Nature Publishing Group},
  issn = {2041-1723},
  doi = {10.1038/s41467-023-44114-0},
  urldate = {2024-10-17},
  copyright = {2024 The Author(s)},
  langid = {english}
}

@article{vandiepen_quantum_2021,
  title = {Quantum {{Simulation}} of {{Antiferromagnetic Heisenberg Chain}} with {{Gate-Defined Quantum Dots}}},
  author = {{van Diepen}, C. J. and Hsiao, T.-K. and Mukhopadhyay, U. and Reichl, C. and Wegscheider, W. and Vandersypen, L. M. K.},
  year = {2021},
  month = nov,
  journal = {Phys. Rev. X},
  volume = {11},
  number = {4},
  pages = {041025},
  publisher = {American Physical Society},
  doi = {10.1103/PhysRevX.11.041025},
  urldate = {2025-08-12}
}

@article{vanloo_photonmediated_2013,
  title = {Photon-{{Mediated Interactions Between Distant Artificial Atoms}}},
  author = {{van Loo}, Arjan F. and Fedorov, Arkady and Lalumi{\`e}re, Kevin and Sanders, Barry C. and Blais, Alexandre and Wallraff, Andreas},
  year = {2013},
  month = dec,
  journal = {Science},
  volume = {342},
  number = {6165},
  pages = {1494--1496},
  publisher = {American Association for the Advancement of Science},
  doi = {10.1126/science.1244324},
  urldate = {2025-08-18}
}

@article{vermersch_implementation_2016,
  title = {Implementation of Chiral Quantum Optics with {{Rydberg}} and Trapped-Ion Setups},
  author = {Vermersch, Beno{\^i}t and Ramos, Tom{\'a}s and Hauke, Philipp and Zoller, Peter},
  year = {2016},
  month = jun,
  journal = {Phys. Rev. A},
  volume = {93},
  number = {6},
  pages = {063830},
  publisher = {American Physical Society},
  doi = {10.1103/PhysRevA.93.063830},
  urldate = {2025-08-24}
}

@article{vlasiuk_cavityinduced_2023,
  title = {Cavity-Induced Charge Transfer in Periodic Systems: {{Length-gauge}} Formalism},
  shorttitle = {Cavity-Induced Charge Transfer in Periodic Systems},
  author = {Vlasiuk, Ekaterina and Kozin, Valerii K. and Klinovaja, Jelena and Loss, Daniel and Iorsh, Ivan V. and Tokatly, Ilya V.},
  year = {2023},
  month = aug,
  journal = {Phys. Rev. B},
  volume = {108},
  number = {8},
  pages = {085410},
  publisher = {American Physical Society},
  doi = {10.1103/PhysRevB.108.085410},
  urldate = {2025-05-02}
}

@article{wang_experimental_2022,
  title = {Experimental Realization of an Extended {{Fermi-Hubbard}} Model Using a {{2D}} Lattice of Dopant-Based Quantum Dots},
  author = {Wang, Xiqiao and Khatami, Ehsan and Fei, Fan and Wyrick, Jonathan and Namboodiri, Pradeep and Kashid, Ranjit and Rigosi, Albert F. and Bryant, Garnett and Silver, Richard},
  year = {2022},
  month = nov,
  journal = {Nat Commun},
  volume = {13},
  number = {1},
  pages = {6824},
  publisher = {Nature Publishing Group},
  issn = {2041-1723},
  doi = {10.1038/s41467-022-34220-w},
  urldate = {2025-08-12},
  copyright = {2022 This is a U.S. Government work and not under copyright protection in the US; foreign copyright protection may apply},
  langid = {english}
}

@article{wang_probing_2023,
  title = {Probing Resonating Valence Bonds on a Programmable Germanium Quantum Simulator},
  author = {Wang, Chien-An and D{\'e}prez, Corentin and Tidjani, Hanifa and Lawrie, William I. L. and Hendrickx, Nico W. and Sammak, Amir and Scappucci, Giordano and Veldhorst, Menno},
  year = {2023},
  month = jun,
  journal = {npj Quantum Inf},
  volume = {9},
  number = {1},
  pages = {58},
  publisher = {Nature Publishing Group},
  issn = {2056-6387},
  doi = {10.1038/s41534-023-00727-3},
  urldate = {2025-08-12},
  copyright = {2023 The Author(s)},
  langid = {english}
}

@misc{wauters_engineering_2024,
  title = {Engineering a {{Josephson}} Junction Chain for the Simulation of the Clock Model},
  author = {Wauters, Matteo M. and Maffi, Lorenzo and Burrello, Michele},
  year = {2024},
  month = aug,
  number = {arXiv:2408.14549},
  eprint = {2408.14549},
  primaryclass = {cond-mat, physics:quant-ph},
  publisher = {arXiv},
  urldate = {2024-10-11},
  archiveprefix = {arXiv}
}

@article{wu_potts_1982,
  title = {The {{Potts}} Model},
  author = {Wu, F. Y.},
  year = {1982},
  month = jan,
  journal = {Rev. Mod. Phys.},
  volume = {54},
  number = {1},
  pages = {235--268},
  publisher = {American Physical Society},
  doi = {10.1103/RevModPhys.54.235},
  urldate = {2024-10-24}
}

@article{yoshihara_superconducting_2017,
  title = {Superconducting Qubit--Oscillator Circuit beyond the Ultrastrong-Coupling Regime},
  author = {Yoshihara, Fumiki and Fuse, Tomoko and Ashhab, Sahel and Kakuyanagi, Kosuke and Saito, Shiro and Semba, Kouichi},
  year = {2017},
  month = jan,
  journal = {Nature Phys},
  volume = {13},
  number = {1},
  pages = {44--47},
  publisher = {Nature Publishing Group},
  issn = {1745-2481},
  doi = {10.1038/nphys3906},
  urldate = {2024-08-28},
  copyright = {2016 Springer Nature Limited},
  langid = {english}
}

@article{zhang_z_n_2014,
  title = {\${{Z}}\_{{N}}\$ Symmetric Chiral {{Rabi}} Model: A New \${{N}}\$-Level System},
  shorttitle = {\${{Z}}\_{{N}}\$ Symmetric Chiral {{Rabi}} Model},
  author = {Zhang, Yao-Zhong},
  year = {2014},
  month = aug,
  journal = {Annals of Physics},
  volume = {347},
  primaryclass = {cond-mat, physics:math-ph, physics:quant-ph},
  pages = {122--129},
  issn = {00034916},
  urldate = {2024-04-15},
  archiveprefix = {arXiv}
}

@article{zhang_observation_2017,
  title = {Observation of a Discrete Time Crystal},
  author = {Zhang, J. and Hess, P. W. and Kyprianidis, A. and Becker, P. and Lee, A. and Smith, J. and Pagano, G. and Potirniche, I.-D. and Potter, A. C. and Vishwanath, A. and Yao, N. Y. and Monroe, C.},
  year = {2017},
  month = mar,
  journal = {Nature},
  volume = {543},
  number = {7644},
  pages = {217--220},
  publisher = {Nature Publishing Group},
  issn = {1476-4687},
  doi = {10.1038/nature21413},
  urldate = {2025-08-12},
  copyright = {2017 Macmillan Publishers Limited, part of Springer Nature. All rights reserved.},
  langid = {english}
}

@book{forschungszentrumjulichgermany_lecture_2024,
  title = {Lecture {{Notes}} on {{Quantum Electrical Circuits}}},
  author = {{Forschungszentrum J{\"u}lich, Germany} and Ciani, Alessandro and DiVincenzo, David P. and {Forschungszentrum J{\"u}lich, Germany} and Terhal, Barbara M. and {Delft University of Technology}},
  year = {2024},
  month = feb,
  publisher = {TU Delft OPEN Publishing},
  doi = {10.59490/tb.85},
  urldate = {2025-09-30},
  isbn = {978-94-6366-815-6},
  langid = {english}
}

@article{KozinMiserevSchottky,
  title = {Schottky anomaly in a cavity-coupled double quantum well},
  author = {Kozin, Valerii K. and Miserev, Dmitry and Loss, Daniel and Klinovaja, Jelena},
  journal = {Phys. Rev. Res.},
  volume = {7},
  issue = {3},
  pages = {033239},
  numpages = {9},
  year = {2025},
  month = {Sep},
  publisher = {American Physical Society},
  doi = {10.1103/43vj-wst3},
  url = {https://link.aps.org/doi/10.1103/43vj-wst3}
}

@article{loss_quantum_1998,
  title = {Quantum Computation with Quantum Dots},
  author = {Loss, Daniel and DiVincenzo, David P.},
  year = 1998,
  month = jan,
  journal = {Physical Review A},
  volume = {57},
  number = {1},
  pages = {120--126},
  publisher = {American Physical Society},
  doi = {10.1103/PhysRevA.57.120},
  urldate = {2024-02-29}
}

@article{ronetti_clock_2021,
  title = {Clock Model and Parafermions in {{Rashba}} Nanowires},
  author = {Ronetti, Flavio and Loss, Daniel and Klinovaja, Jelena},
  year = 2021,
  month = jun,
  journal = {Physical Review B},
  volume = {103},
  number = {23},
  pages = {235410},
  publisher = {American Physical Society},
  doi = {10.1103/PhysRevB.103.235410},
  urldate = {2022-12-19}
}

@article{calzona_$mathbbz_4$_2018a,
  title = {\$\textbraceleft\textbackslash mathbb\textbraceleft{{Z}}\textbraceright\textbraceright\_\textbraceleft 4\textbraceright\$ Parafermions in One-Dimensional Fermionic Lattices},
  author = {Calzona, Alessio and Meng, Tobias and Sassetti, Maura and Schmidt, Thomas L.},
  year = 2018,
  month = nov,
  journal = {Physical Review B},
  volume = {98},
  number = {20},
  pages = {201110},
  publisher = {American Physical Society},
  doi = {10.1103/PhysRevB.98.201110},
  urldate = {2026-04-10}
}

@article{mazza_nontopological_2018,
  title = {Nontopological Parafermions in a One-Dimensional Fermionic Model with Even Multiplet Pairing},
  author = {Mazza, Leonardo and Iemini, Fernando and Dalmonte, Marcello and Mora, Christophe},
  year = 2018,
  month = nov,
  journal = {Physical Review B},
  volume = {98},
  number = {20},
  pages = {201109},
  publisher = {American Physical Society},
  doi = {10.1103/PhysRevB.98.201109},
  urldate = {2026-04-10}
}

\end{document}